%% file: dopplercc.tex
\documentclass[conference]{IEEEtran}
\IEEEoverridecommandlockouts
\usepackage{cite}
\usepackage{amsmath,amssymb,amsfonts}
\usepackage[nolist]{acronym}
\usepackage{algorithmic}
\usepackage{hyperref}
\usepackage{scalerel}
\usepackage{graphicx}
\usepackage{textcomp}
\usepackage{xcolor}
\usepackage{tikz}
\usepackage{nicefrac}
\usepackage{circuitikz}

\newdimen\bpt
\def\mobile#1{\leavevmode 
   \bpt=#1bp \hbox to7\bpt{\kern1\bpt \lower1\bpt\vbox to12\bpt{}%
      \pdfliteral{q #1 0 0 #1 0 0 cm 1 j 2 w 0 0 5 10 re B 
         1 g 1 G  1 w .3 1.8 4.4 7 re B 
         1.5 w 2.5 .2 0 .1 re B .3 w 1.7 10 1.6 0 re B Q}%
      \hss}}

\usetikzlibrary{calc}
\usetikzlibrary{positioning}
\usepackage{pgfplots}
\usepgfplotslibrary{groupplots}
\usepgfplotslibrary{colorbrewer}
\pgfplotsset{compat = 1.18}

\makeatletter
\let\MYcaption\@makecaption
\makeatother

\usepackage[font=footnotesize]{subcaption}

\makeatletter
\let\@makecaption\MYcaption
\makeatother

\DeclareMathOperator*{\argmin}{arg\,min}

\usetikzlibrary{svg.path}

\def\BibTeX{{\rm B\kern-.05em{\sc i\kern-.025em b}\kern-.08em
    T\kern-.1667em\lower.7ex\hbox{E}\kern-.125emX}}

\definecolor{orcidlogocol}{HTML}{A6CE39}
\tikzset{
  orcidlogo/.pic={
    \fill[orcidlogocol] svg{M256,128c0,70.7-57.3,128-128,128C57.3,256,0,198.7,0,128C0,57.3,57.3,0,128,0C198.7,0,256,57.3,256,128z};
    \fill[white] svg{M86.3,186.2H70.9V79.1h15.4v48.4V186.2z}
                 svg{M108.9,79.1h41.6c39.6,0,57,28.3,57,53.6c0,27.5-21.5,53.6-56.8,53.6h-41.8V79.1z M124.3,172.4h24.5c34.9,0,42.9-26.5,42.9-39.7c0-21.5-13.7-39.7-43.7-39.7h-23.7V172.4z}
                 svg{M88.7,56.8c0,5.5-4.5,10.1-10.1,10.1c-5.6,0-10.1-4.6-10.1-10.1c0-5.6,4.5-10.1,10.1-10.1C84.2,46.7,88.7,51.3,88.7,56.8z};
  }
}

\newcommand\orcidicon[1]{\href{https://orcid.org/#1}{\mbox{\scalerel*{
\begin{tikzpicture}[yscale=-1,transform shape]
\pic{orcidlogo};
\end{tikzpicture}
}{|}}}}

\input{acronyms.tex}
\input{corporate_colors.tex}

\begin{document}

\title{Leveraging the Doppler Effect for Channel Charting
\thanks{This work is supported by the German Federal Ministry of Education and Research (BMBF) within the projects Open6GHub (grant no. 16KISK019) and KOMSENS-6G (grant no. 16KISK113).}}

\author{\IEEEauthorblockN{Florian Euchner\textsuperscript{\orcidicon{0000-0002-8090-1188}}, Phillip Stephan\textsuperscript{\orcidicon{0009-0007-4036-668X}}, Stephan ten Brink\textsuperscript{\orcidicon{0000-0003-1502-2571}} \\}

\IEEEauthorblockA{
Institute of Telecommunications, Pfaffenwaldring 47, University of  Stuttgart, 70569 Stuttgart, Germany \\ \{euchner,stephan,tenbrink\}@inue.uni-stuttgart.de
}
}

\maketitle

\begin{abstract}
Channel Charting is a dimensionality reduction technique that reconstructs a map of the radio environment from similarity relationships found in channel state information.
Distances in the channel chart are often computed based on some dissimilarity metric, which can be derived from angular-domain information, channel impulse responses, measured phase differences or simply timestamps.
Using such information implicitly makes strong assumptions about the level of phase and time synchronization between base station antennas or assumes approximately constant transmitter velocity.
Many practical systems, however, may not provide phase and time synchronization and single-antenna base stations may not even have angular-domain information.
We propose a Doppler effect-based loss function for Channel Charting that only requires frequency synchronization between spatially distributed base station antennas, which is a much weaker assumption.
We use a dataset measured in an indoor environment to demonstrate that the proposed method is practically feasible with just four base station antennas, that it produces a channel chart that is suitable for localization in the global coordinate frame and that it outperforms other state-of-the-art methods under the given limitations.
\end{abstract}

\begin{IEEEkeywords}
Channel Charting, Doppler effect, localization
\end{IEEEkeywords}

\section{Introduction}
Channel Charting \cite{studer_cc} is a self-supervised dimensionality reduction method that learns the \ac{FCF}, a mapping from high-dimensional \ac{CSI} to a low-dimensional representation, the so-called channel chart, which can be interpreted as a map of the radio environment.
It relies on similarity relationships in measured \ac{CSI}, thus exploiting the multipath characteristics of the propagation environment.
Initially, Channel Charting was proposed primarily as a technique for for \emph{relative} localization, i.e., only achieving local spatial consistency, but additional work \cite{fraunhofer_cc, stephan2023angle} suggested that even the \emph{global} geometry could be preserved and that Channel Charting in absolute global coordinates is possible as long as the location of \ac{BS} antennas is known \cite{pihlajasalo2020absolute, taner2023channel, asilomar2023}.
Previous work on Channel Charting in global coordinates assumed a large number of time- and phase-synchronized antennas, with a total of 80 \ac{BS} antennas in \cite{pihlajasalo2020absolute}, and 32 \ac{BS} antennas in \cite{taner2023channel} and \cite{asilomar2023}.
Here, we implement Channel Charting in global coordinates with only 4 \ac{BS} antennas and without time or phase synchronization, by leveraging Doppler effect-induced phase changes\footnote{Partial source code and the dataset for this paper are publicly available at \url{ https://github.com/Jeija/Doppler-Effect-ChannelCharting/}}.

\begin{figure*}
    \centering
    \begin{subfigure}[b]{0.30\textwidth}
        \centering
        \includegraphics[width=0.8\textwidth, trim = 30 150 30 0, clip]{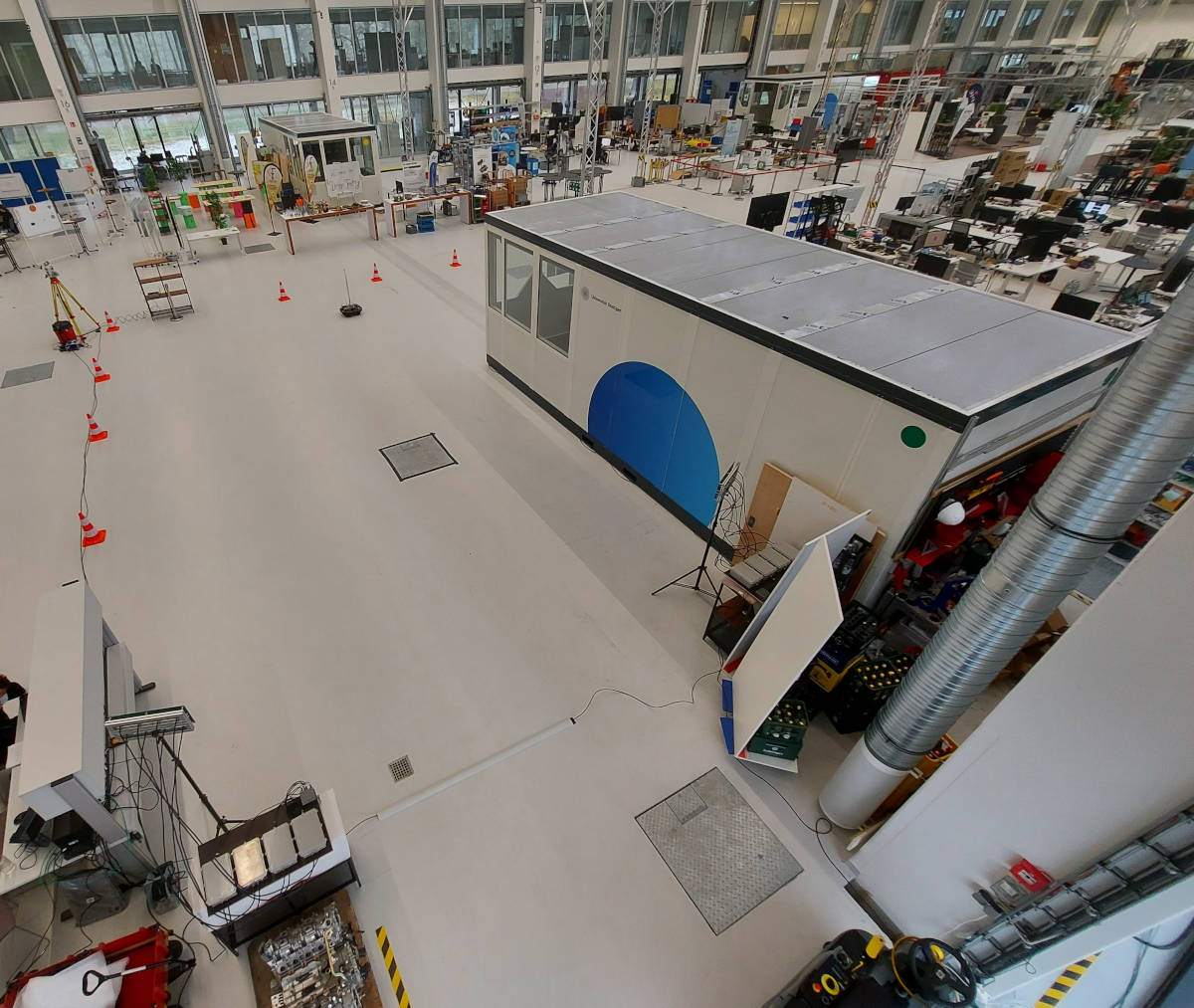}
        \vspace{0.75cm}
        \caption{}
    \end{subfigure}
    \begin{subfigure}[b]{0.3\textwidth}
        \centering
        \begin{tikzpicture}
            \begin{axis}[
                width=0.729\columnwidth,
                height=0.6\columnwidth,
                scale only axis,
                xmin=-15.5,
                xmax=6.1,
                ymin=-18.06,
                ymax=-1.5,
                xlabel = {Coordinate $x_1 ~ [\mathrm{m}]$},
                ylabel = {Coordinate $x_2 ~ [\mathrm{m}]$},
                ylabel shift = -8 pt,
                xlabel shift = -4 pt,
                xtick={-10, -6, -2, 2}
            ]
                \addplot[thick,blue] graphics[xmin=-14.5,ymin=-17.06,xmax=4.1,ymax=-1.5] {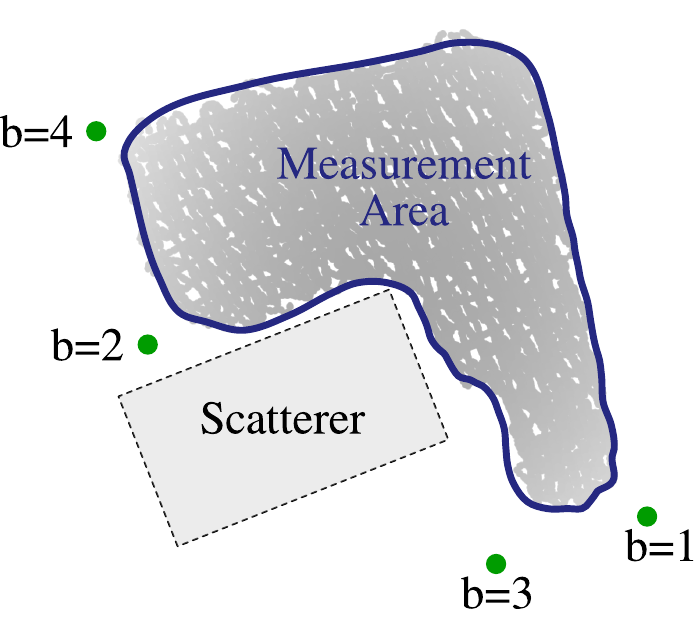};
            \end{axis}
        \end{tikzpicture}
        \vspace{-0.2cm}
        \caption{}
        \label{fig:labelled-area}
    \end{subfigure}
    \begin{subfigure}[b]{0.3\textwidth}
        \centering
        \begin{tikzpicture}
            \begin{axis}[
                width=0.6\columnwidth,
                height=0.6\columnwidth,
                scale only axis,
                xmin=-12.5,
                xmax=2.5,
                ymin=-14.5,
                ymax=-1.5,
                xlabel = {Coordinate $x_1 ~ [\mathrm{m}]$},
                ylabel = {Coordinate $x_2 ~ [\mathrm{m}]$},
                ylabel shift = -8 pt,
                xlabel shift = -4 pt,
                xtick={-10, -6, -2, 2}
            ]
                \addplot[thick,blue] graphics[xmin=-12.5,ymin=-14.5,xmax=2.5,ymax=-1.5] {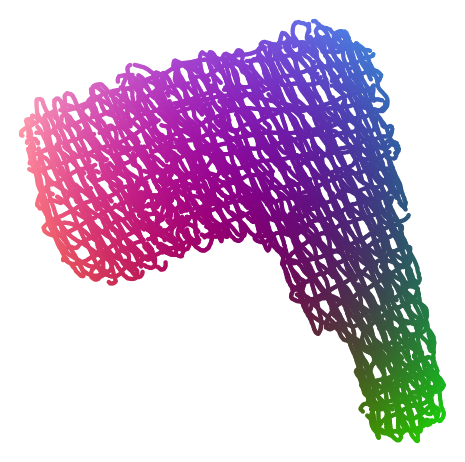};
            \end{axis}
        \end{tikzpicture}
        \vspace{-0.2cm}
        \caption{}
        \label{fig:groundtruth-map}
    \end{subfigure}
    \vspace{-0.2cm}
    \caption{Information about the environment the dataset was measured in: The figure shows (a) a photograph of the environment, (b) a top view map and (c) a scatter plot of colorized ``ground truth'' positions (only used for evaluation, not training) of datapoints in $\mathcal S_\mathrm{train}$.}
    \label{fig:industrial_environment}
    \vspace{-0.3cm}
\end{figure*}

The classical problem of source localization and velocity estimation for an active transmitter based on Doppler shift measurements from a spatially distributed sensor network is well-known and has been thoroughly investigated, e.g., in \mbox{\cite{chan, shames}}.
Their work uses Doppler shift measurements to not only estimate the transmitter's velocity, but also its location and \ac{CFO}.
However, classical solutions to this problem, which optimize a cost function parametrized on individual frequency shift measurements, have several shortcomings:
For example, \cite{chan} assumes a nonmaneuvering transmitter, i.e., a constant transmitter velocity over a certain timespan.
The approach in \cite{shames}, on the other hand, needs Doppler shift measurements at five or more locations to solve for the unknown variables (\ac{CFO}, position and velocity in two dimensions), even in the noiseless case.
In contrast to that, our proposed Channel Charting-based approach can locate users with fewer \ac{BS} antennas.
It learns the \ac{FCF}, implemented as a neural network, from a large number of observations and then uses that \ac{FCF} for localization.
Whereas the classical approaches only work \emph{while} the transmitter is moving, our Channel Charting-based technique, once trained on moving transmitters, will continue to function for static transmitters.


\section{Dataset and System Model}
\label{sec:dataset}
In our system model, we have a single transmitter, called \ac{UE}, and several frequency-synchronized and spatially distributed receiver antennas (single-antenna \acp{BS}).
In practice, frequency synchronization among \ac{BS} antennas can be achieved by distributing a reference clock signal, by exchanging over-the-air synchronization messages, or by relying on time sources like \acp{GPSDO}.
A central challenge in Doppler effect-based radio localization is the unknown and variable transmit frequency.
While the carrier frequency should ideally be well-defined, it is usually derived from a crystal oscillator, which can exhibit large offsets and drifts.
A typical crystal oscillator can have a frequency offset of $\pm20\,\mathrm{ppm}$, corresponding to a \ac{CFO}-induced frequency shift of $f_\mathrm{CFO} = \pm20\,\mathrm{kHz}$ even at a low carrier frequency of $1\,\mathrm{GHz}$.
At that frequency, moving at a velocity of $1\,\frac{\mathrm m}{\mathrm s}$ leads to a Doppler shift of just around $\pm 3.3\,\mathrm{Hz}$!
A Doppler effect-based localization system must hence be robust against \ac{CFO}.

The considered indoor factory environment dataset, known as \emph{dichasus-cf0x}, was already used in previous work on Channel Charting \cite{stephan2023angle, stahlke2023velocity, taner2023streaming, asilomar2023}.
The dataset was generated by \emph{\ac{DICHASUS}}, our distributed \ac{mMIMO} channel sounder, whose architecture is described in \cite{dichasus2021}.
In brief, \ac{DICHASUS} measures the channel between the \ac{UE} and all \ac{BS} antennas.
\ac{DICHASUS} provides large datasets containing frequency-domain channel coefficients, alongside timestamps and accurate position information.
For \emph{dichasus-cf0x}, $B = 4$ receiver antenna arrays with $2\times4$ antennas each are set up in a factory hall, and $N_\mathrm{sub} = 1024$ \ac{OFDM} channel coefficients are measured at a carrier frequency of $1.272\,\mathrm{GHz}$ and with a total bandwidth of $50\,\mathrm{MHz}$.
The $B = 4$ antennas are known to be located at $\mathbf z^{(b)} \in \mathbb R^3$.
Here, we only use \ac{CSI} from one antenna per array, and we arbitrarily pick the third antenna in the first row of each array.
Hence, the resulting \ac{CSI} can be represented as a matrix $\mathbf H^{(l)} \in \mathbb C^{B \times N_\mathrm{sub}}$ containing complex-valued channel coefficients of all $B = 4$ \ac{BS} antennas and all $N_\mathrm{sub}$ subcarriers for time instance $l$.
$\mathcal F^{-1} \left\{\mathbf H^{(l)} \right\}$ denotes the inverse discrete Fourier transform of the \ac{CSI} matrix along the frequency axis, i.e., the antenna-specific \acp{CIR}.

The transmitter is a dipole antenna which is mounted on top of a robot, traveling at a median speed of approximately $0.25\,\frac{\mathrm m}{\mathrm s}$ along a set of trajectories inside an L-shaped area, with an overall bounding box size of approximately $14\,\mathrm{m} \times 14\,\mathrm{m}$.
A prism is attached to the tip of the transmit antenna and tracked with millimeter-level precision by a tachymeter, providing ``ground truth'' positions $\mathbf x^{(l)}$.
In all localization tasks, we assume that the height of the transmitter, $x_3^{(l)}$, is known and constant, which simplifies the problem to two-dimensional localization.
In addition to \ac{CSI} and reference positions, timestamps $t^{(l)}$ as well as instantaneous, \ac{BS} antenna-specific frequency offset measurements $\mathbf f^{(l)} \in \mathbb R^B$ are collected.
Thus, the complete dataset can be represented as the following set containing a total of $L$ datapoints:
\[
    \text{Dataset}: \mathcal S = \left\{ \left(\mathbf H^{(l)}, \mathbf x^{(l)}, t^{(l)}, \mathbf f^{(l)} \right) \right\}_{l = 1, \ldots, L}
\]

\begin{figure}
    \centering
    \begin{subfigure}[b]{0.4\columnwidth}
        \begin{circuitikz}
            \node (bs) [dinantenna] at (0, 0) {};
            \node [below = 0cm of bs] {BS};
            \node (ue) [align = center] at (2, 0.2) {\mobile{1.5}\\UE};
            \draw [dashed, latex-latex] (bs.east) -- (ue.west |- bs.east) node[midway, above] {$d(t_2)$};
            \node (circ) [circle, inner sep = 1pt, fill, red] at ($(ue.center) + (0, 0.2)$) {};
            \draw [thick, red, -latex] (circ) -- +(0.8, 0) node[midway, above] {v};
        \end{circuitikz}
        \vspace{0.2cm}
    \end{subfigure}
    \begin{subfigure}[b]{0.55\columnwidth}
        \begin{tikzpicture}[
          declare function={func(\x,\a) = 2 * 3.1415 * mod(\x,\a)/\a - 3.1415;} 
        ]
            \begin{axis}[
                axis x line = middle,
                axis y line = middle,
                samples = 401,
                ymax=3.5,
                ymin=-3.5,
                domain=0:2.7,
                xmin = 0,
                xmax = 3,
                no marks,
                width = \textwidth,
                height = 0.6\textwidth,
                ytick = {-3.1415, 0, 3.1415},
                yticklabels = {$-\pi$, 0, $\pi$},
                xtick = {1, 2},
                xticklabel style = {fill = white, text opacity = 1, fill opacity = 0.7, font=\footnotesize},
                set layers = axis on top,
                xlabel = {$t \left[ \frac{\lambda}{v} \right]$},
                x label style={at={(axis description cs:1,0.5)}, anchor=north},
                ylabel = $\varphi_1(t_2) ~ \mathrm{mod} ~ 2\pi$,
                y label style={at={(axis description cs:0,1)}, anchor=south}
            ]
                \addplot [red!80!white] {-func(100 + 2000 * \x, 200)};
            \end{axis}
        \end{tikzpicture}
    \end{subfigure}
    \vspace{-0.3cm}
    \caption{Single BS antenna, unsynchronized UE moving at velocity $v$ away from BS: Received uplink phase is affected by both CFO and Doppler shift.}
    \label{fig:doppler_singlebs}
    \vspace{-0.3cm}
\end{figure}
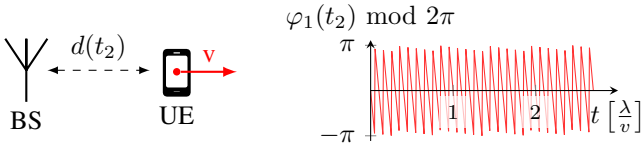

For evaluation purposes, the dataset $\mathcal S$ is split into a training set $\mathcal S_\mathrm{train}$ and a test set $\mathcal S_\mathrm{test}$ containing $|\mathcal S_\mathrm{train}| = |\mathcal S_\mathrm{test}| = 20851$ datapoints each.
A photo and a top view map of the environment are shown in Fig.~\ref{fig:industrial_environment}.
The true datapoint positions $\mathbf x^{(l)}$ are shown in Fig.~\ref{fig:groundtruth-map}.
The points have been colorized and the datapoints will retain their color even as the FCF maps them to a position in the channel chart.
This allows for a visual evaluation of the generated chart: If the global topology is preserved, a similar color gradient should appear in the chart.

Thanks to an elaborate over-the-air synchronization technique \cite{dichasus2021, euchner2022geometry}, DICHASUS achieves frequency, time and phase coherence across all spatially distributed antennas.
However, to support the claim that the Doppler effect-based localization technique works even without phase and time synchronization, we remove the fine time synchronization and randomize the initial phase for the feature vectors provided to the neural network.
Under these circumstances, \ac{ToA}-based positioning would not work.
Furthermore, with only one antenna at each \ac{BS} site, angular-domain information is not available either, rendering \ac{AoA}-based positioning infeasible.
Common Channel Charting techniques and dissimilarity metrics \cite{studer_cc, fraunhofer_cc, stephan2023angle} equally fail due to making assumptions about synchronization.
We want to point out that carrier and sampling frequency synchronization implies phase and time synchronization down to constant offsets over short time intervals.
However, we acknowledge that the frequency synchronization may be imperfect, i.e., phases and sampling times may drift over longer time intervals.
Furthermore, \ac{PLL}-based frequency synthesizers are another well-known source of output phase ambiguity \cite{brown2005method}, causing loss of time and phase synchronization when the \ac{PLL} re-locks.

\section{Doppler Effect-Based Loss Function}
\label{sec:loss}
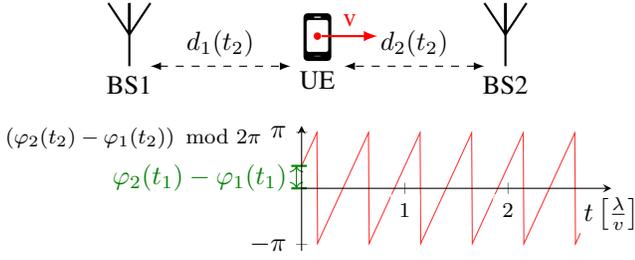
\begin{figure}
    \centering
    \begin{subfigure}[b]{0.9\columnwidth}
        \centering
        \begin{circuitikz}
            \node (bs1) [dinantenna] at (0, 0) {};
            \node [below = 0cm of bs1] {BS1};
            \node (bs2) [dinantenna] at (5, 0) {};
            \node [below = 0cm of bs2] {BS2};
            \node (ue) [align = center] at (2.5, 0.2) {\mobile{1.5}\\UE};
            \draw [dashed, latex-latex] (bs1.south east) -- (ue.west |- bs1.south east) node[midway, above] {$d_1(t_2)$};
            \draw [dashed, latex-latex] (bs2.south west) -- (ue.east |- bs2.south west) node[midway, above] {$d_2(t_2)$};
            \node (circ) [circle, inner sep = 1pt, fill, red] at ($(ue.center) + (0, 0.2)$) {};
            \draw [thick, red, -latex] (circ) -- +(0.8, 0) node[midway, above] {v};
        \end{circuitikz}
        \vspace{0.2cm}
    \end{subfigure}
    \begin{subfigure}[b]{0.99\columnwidth}
        \centering
        \begin{tikzpicture}[
          declare function={func(\x,\a) = 2 * 3.1415 * mod(\x,\a)/\a - 3.1415;} 
        ]
            \begin{axis}[
                axis x line = middle,
                axis y line = middle,
                samples = 401,
                ymax=3.5,
                ymin=-3.5,
                domain=0:2.7,
                xmin = 0,
                xmax = 3,
                no marks,
                width = 0.65\textwidth,
                height = 0.37\textwidth,
                ytick = {-3.1415, 0, 3.1415},
                yticklabels = {$-\pi$, 0, $\pi$},
                xtick = {1, 2},
                xticklabel style = {fill = white, text opacity = 1, fill opacity = 0.7, font=\footnotesize},
                set layers = axis on top,
                xlabel = {$t \left[ \frac{\lambda}{v} \right]$},
                x label style={at={(axis description cs:1,0.5)}, anchor=north},
                ylabel = $\left(\varphi_2(t_2) - \varphi_1(t_2)\right) ~ \mathrm{mod} ~ 2\pi$,
                y label style={at={(axis description cs:-0.1,0.9)}, anchor=east, font=\footnotesize},
                clip = false
            ]
                \addplot [red!80!white] {func(700 + 2000 * \x, 1000)};
                \draw [thick, green!50!black] (axis cs:-0.1,1.257) -- (axis cs:0.05,1.257);
                \draw [thick, green!50!black] (axis cs:-0.1,0) -- (axis cs:0.05,0);
                \draw [green!50!black, <->] (axis cs:-0.05,0) -- (axis cs:-0.05,1.257) node[midway, left] {$\varphi_{2}(t_1) - \varphi_{1}(t_1)$};
            \end{axis}
        \end{tikzpicture}
    \end{subfigure}
    \vspace{-0.4cm}
    \caption{Two BS antennas, UE moving at velocity $v$ between BS antennas. The change in differential uplink phases, caused by the Doppler effect, is now unaffected by CFO. $\lambda$ denotes the wavelength. The figure assumes $t_1 = 0$.}
    \label{fig:doppler_twobs}
    \vspace{-0.3cm}
\end{figure}

In the following, we derive a loss function for training the \ac{NN} that implements the \ac{FCF}.
Since our loss function operates on pairs of positions, it is particularly suited for training the \ac{FCF} embedded into a Siamese \ac{NN} configuration.

\subsection{Introduction: Doppler Effect and Cumulative Phases}
The causes for aforementioned observed frequency offsets $\mathbf f \in \mathbb R^B$, stored the dataset, are twofold:
The \ac{CFO} between \ac{UE} and \ac{BS} antennas $f_\mathrm{CFO}$, and the antenna-specific Doppler shift $f_{\mathrm{Dop}, b}$, $b \in \{ 1, \ldots, B \}$ where usually $|f_\mathrm{CFO}| \gg |f_{\mathrm{Dop}, b}|$.
Assuming a free space channel model, the frequency offset caused by the movement of the \ac{UE} in space leads to a change in the received phase at the \ac{BS} $b$.
In a continuous-time model, the received phase at time $t_2$ can be described as
\begin{equation}
    \begin{split}
        \varphi_b(t_2) &= \int_{t_1}^{t_2} 2 \pi f_b(t) \, \mathrm dt = \int_{t_1}^{t} 2 \pi \left(f_\mathrm{CFO}(t) + f_\mathrm{Dop, b}(t)\right) \, \mathrm dt \\
        &= \varphi_\mathrm{b}(t_1) + \varphi_\mathrm{CFO}(t_2) + \varphi_\mathrm{Dop, b}(t_2),
    \end{split}\raisetag{0.5\baselineskip}
    \label{eq:integration}
\end{equation}
where $\varphi_b(t_1)$ is the initial phase for \ac{BS} $b$ at time $t_1$, $\varphi_\mathrm{CFO}(t_2)$ denotes the \ac{CFO}-induced phase change and $\varphi_\mathrm{Dop, b}(t_2)$ denotes the \ac{BS} antenna-specific phase change caused by the Doppler effect.
We always assume $t_2 \geq t_1$ without loss of generality.
Due to the large and unknown \ac{CFO}, $\varphi_b(t_2)$ by itself is not useful for localization or velocity estimation, as illustrated in Fig.~\ref{fig:doppler_singlebs}:
The phase drift due to \ac{CFO} far outpaces the phase change caused by movement.
However, the \emph{difference} in phase changes $\varphi_b(t_2)$ between two \acp{BS} $b_1$ and $b_2$ is unaffected by \ac{CFO}, since both \ac{BS} antennas, being frequency-synchronized, observe the same \ac{CFO}:
\begin{equation}
    \begin{split}
        &\varphi_{b_2}(t_2) - \varphi_{b_1}(t_2) \\
        =& \varphi_\mathrm{b_2}(t_1) - \varphi_\mathrm{b_1}(t_1) + \varphi_\mathrm{Dop, b_2}(t_2) - \varphi_\mathrm{Dop, b_1}(t_2).        
    \end{split}
    \label{eq:diff_phases}
\end{equation}

This is visualized in the one-dimensional model in Fig.~\ref{fig:doppler_twobs}, where the rate of change in the received phase differences (i.e., the angular frequency) could be used to estimate the velocity $v$ of the \ac{UE}.
We realize that $\frac{\lambda}{2\pi} (\varphi_\mathrm{Dop, b}(t_2) - \varphi_\mathrm{Dop, b}(t_1))$ is the change in distance between \ac{UE} and \ac{BS} antenna $b$ when a Doppler shift-induced phase change of $\varphi_\mathrm{Dop, b}(t_2) - \varphi_\mathrm{Dop, b}(t_1)$ is measured.
By solving Eq.~(\ref{eq:diff_phases}) for $\varphi_\mathrm{Dop, b_2}(t_2) - \varphi_\mathrm{Dop, b_1}(t_2)$ and assuming unwrapped phases, we can obtain an expression relating the phase change to the displacement of the \ac{UE} in the time interval $[t_1, t_2]$:
\begin{equation}
    \begin{split}
        &\varphi_\mathrm{Dop, b_2}(t_2) - \varphi_\mathrm{Dop, b_1}(t_2) \\
        =& \left(\varphi_{b_2}(t_2) - \varphi_{b_1}(t_2)\right) - \left(\varphi_\mathrm{b_2}(t_1) - \varphi_\mathrm{b_1}(t_1)\right) \\
        =\frac{2 \pi}{\lambda} &\left[ \left(d_{b_2}(t_2) - d_{b_1}(t_2) \right) - \left( d_{b_2}(t_1) - d_{b_1}(t_1) \right) \right],
    \end{split}\raisetag{3\baselineskip}
    \label{eq:diff_diff}
\end{equation}
where $d_b(t) = \lVert \hat{\mathbf x}(t) - \mathbf z^{(b)} \rVert_2$ denotes the spatial distance between \ac{BS} $b$ and supposed \ac{UE} location $\hat{\mathbf x}(t)$ at time $t$.

\subsection{In Discrete Time, Comparing Phases}
\begin{figure*}
    \centering
    \begin{subfigure}[t]{0.24\textwidth}
        \centering
        \begin{tikzpicture}
            \begin{axis}[
                width=0.65\columnwidth,
                height=0.65\columnwidth,
                scale only axis,
                enlargelimits=false,
                axis on top,
                xlabel = {Coordinate $\hat x_1$},
                ylabel = {Coordinate $\hat x_2$},
                ylabel shift = -8 pt,
                xlabel shift = -4 pt,
                xmin = -12, xmax = 3,
                ymin = -15, ymax = 0
            ]
                \addplot[thick,blue] graphics[xmin=-12, xmax=3, ymin=-15, ymax=0] {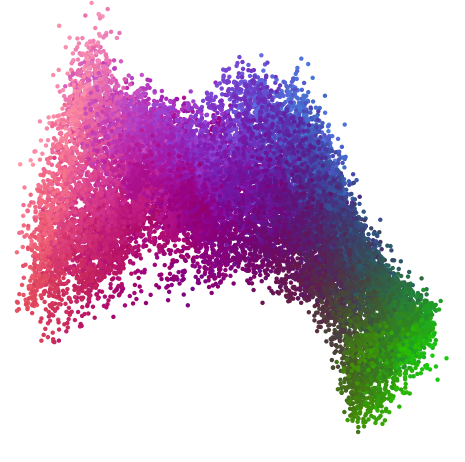};
            \end{axis}
        \end{tikzpicture}
        \vspace{-0.1cm}
        \caption{Channel Chart: $T_\mathrm{opt} \circ C_{\theta, \mathrm{Dop}}$}
        \label{fig:cc_tc_dop}
    \end{subfigure}
    \begin{subfigure}[t]{0.24\textwidth}
        \centering
        \begin{tikzpicture}
            \begin{axis}[
                width=0.65\columnwidth,
                height=0.65\columnwidth,
                scale only axis,
                enlargelimits=false,
                axis on top,
                xlabel = {Coordinate $\hat x_1$},
                ylabel = {Coordinate $\hat x_2$},
                ylabel shift = -8 pt,
                xlabel shift = -4 pt,
                xmin = -12, xmax = 3,
                ymin = -15, ymax = 0
            ]
                \addplot[thick,blue] graphics[xmin=-12, xmax=3, ymin=-15, ymax=0] {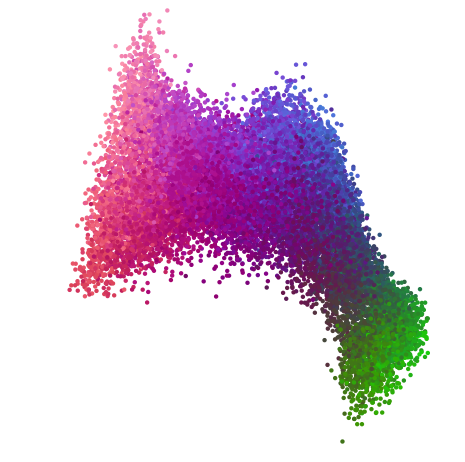};
            \end{axis}
        \end{tikzpicture}
        \vspace{-0.1cm}
        \caption{Channel Chart: $C_{\theta, \mathrm{Dop}}$}
        \label{fig:cc_dop}
    \end{subfigure}
    \begin{subfigure}[t]{0.24\textwidth}
        \centering
        \begin{tikzpicture}
            \begin{axis}[
                width=0.65\columnwidth,
                height=0.65\columnwidth,
                scale only axis,
                enlargelimits=false,
                axis on top,
                xlabel = {Coordinate $\hat x_1$},
                ylabel = {Coordinate $\hat x_2$},
                ylabel shift = -8 pt,
                xlabel shift = -4 pt,
                xmin = -12, xmax = 3,
                ymin = -15, ymax = 0
            ]
                \addplot[thick,blue] graphics[xmin=-12, xmax=3, ymin=-15, ymax=0] {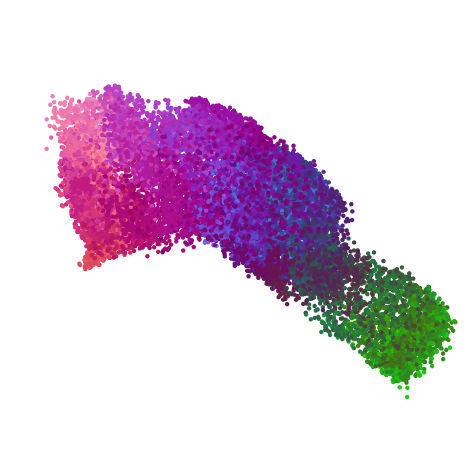};
            \end{axis}
        \end{tikzpicture}
        \vspace{-0.1cm}
        \caption{Channel Chart: $T_\mathrm{opt} \circ C_{\theta, \mathrm{fuse}}$}
        \label{fig:cc_tc_fuse}
    \end{subfigure}
    \begin{subfigure}[t]{0.24\textwidth}
        \centering
        \begin{tikzpicture}
            \begin{axis}[
                width=0.65\columnwidth,
                height=0.65\columnwidth,
                scale only axis,
                enlargelimits=false,
                axis on top,
                xlabel = {Coordinate $\hat x_1$},
                ylabel = {Coordinate $\hat x_2$},
                ylabel shift = -8 pt,
                xlabel shift = -4 pt,
                xmin = -12, xmax = 3,
                ymin = -15, ymax = 0
            ]
                \addplot[thick,blue] graphics[xmin=-12, xmax=3, ymin=-15, ymax=0] {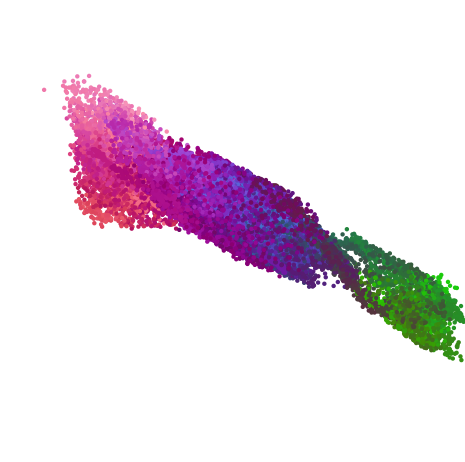};
            \end{axis}
        \end{tikzpicture}
        \vspace{-0.1cm}
        \caption{Channel Chart: $T_\mathrm{opt} \circ C_{\theta, \mathrm{CIRA}}$}
        \label{fig:cc_tc_cira}
    \end{subfigure}
    \vspace{-0.1cm}
    \caption{Learned channel charts, FCF applied to $S_\mathrm{train}$. Colors of points are preserved from color gradient applied in Fig.~\ref{fig:groundtruth-map}.}
    \label{fig:cc-all}
    \vspace{-0.3cm}
\end{figure*}

Eq.~(\ref{eq:integration}) describes how $\varphi_b(t)$ is computed by integrating the observed frequency offsets over time.
In practice, the frequency offset measurements in our dataset are only available at discrete points in time, so we replace the integration with the corresponding summation taking into account differences in timestamps, and obtain the discrete-time phase measurements $\varphi_b^{(l)}$ at times $t^{(l)}$.
In addition to the instantaneous frequency offset estimates $\mathbf f^{(l)}$ in the dataset, we update the integrated phases $\varphi_b^{(l)}$ based on the phase measurements in the \ac{CSI} matrix $\mathbf H^{(l)}$, by comparing phases predicted based on $\mathbf f^{(l)}$ to phases actually measured in $\mathbf H^{(l)}$.
This, of course, requires sufficiently frequent sampling, which is realistic for practical systems that may also transmit very frequently or continuously.

\subsection{Log-Likelihood Function}
We derive a log-likelihood function for pairs of datapoints $(l_1, l_2)$ by generalizing a discrete-time version of Eq.~(\ref{eq:diff_diff}) to more than two \ac{BS} antennas.
To this end, we assume that we have unwrapped phase measurements $\varphi_{b}^{(l)}$ and we assume an i.i.d. normally distributed error in the phase difference measurements $n_{b_1, b_2}^{(l_1, l_2)} \sim \mathcal N \left(0, \left(\sigma_{b_1, b_2}^{(l_1, l_2)}\right)^2 \right)$, i.e.,
\begin{align*}
    &\left( \varphi_{b_2}^{(l_2)} - \varphi_{b_1}^{(l_2)} \right) - \left( \varphi_{b_2}^{(l_1)} - \varphi_{b_1}^{(l_1)} \right) + n_{b_1, b_2}^{(l_1, l_2)} \\
    = \frac{2 \pi}{\lambda} &\left[\left( d_{b_2}^{(l_2)} - d_{b_1}^{(l_2)} \right) - \left( d_{b_2}^{(l_1)} - d_{b_1}^{(l_1)} \right)\right],
\end{align*}
where $d_b^{(l)} = \lVert \hat{\mathbf x}^{(l)} - \mathbf z^{(b)} \rVert_2$.
We remark that the assumption of an i.i.d. error distribution is flawed, and may be revised in future versions.
We let $f_{b_1, b_2}^{(l_1, l_2)}(\hat{\mathbf x}^{(l_1)}, \hat{\mathbf x}^{(l_2)})$ denote the corresponding likelihood function, based on the Gaussian probability density function of $n_{b_1, b_2}^{(l_1, l_2)}$.
For any pair of datapoints $(l_1, l_2)$, based on the assumption of i.i.d. errors $n_{b_1, b_2}^{(l_1, l_2)}$, we find the overall likelihood for the pair of positions $(\mathbf x^{(l_1)}, \mathbf x^{(l_2)})$ taking into account all pairs of \ac{BS} antennas to be
\[
    f^{(l_1, l_2)} = \prod_{b_1=1}^B \prod_{b_2=1}^{B} f_{b_1, b_2}^{(l_1, l_2)}.
\]

Finally, we find the log-likelihood function (neglecting constant terms) \mbox{$\mathcal L^{(l_1, l_2)} = \log f^{(l_1, l_2)}(\hat{\mathbf x}^{(l_1)}, \hat{\mathbf x}^{(l_2)}) - \mathrm{const.}$} as
\begin{equation}
    \mathcal L^{(l_1, l_2)} = \sum_{b_1 = 1}^B \sum_{b_2 = 1}^B \left(\frac{\Delta \varphi_{b_1, b_2}^{(l_1, l_2)} - \frac{2 \pi}{\lambda} \Delta d_{b_1, b_2}^{(l_1, l_2)}}{\sigma_{b_1, b_2}^{(l_1, l_2)}}\right)^2,
    \label{eq:loglikelihood}
\end{equation}
where we use the abbreviations
\begin{equation}
    \begin{split}
        \Delta \varphi_{b_1, b_2}^{(l_1, l_2)} &= \left( \varphi_{b_2}^{(l_2)} - \varphi_{b_1}^{(l_2)} \right) - \left( \varphi_{b_2}^{(l_1)} - \varphi_{b_1}^{(l_1)} \right) ~ \text{and} \\
        \Delta d_{b_1, b_2}^{(l_1, l_2)} &= \left( d_{b_2}^{(l_2)} - d_{b_1}^{(l_2)} \right) - \left( d_{b_2}^{(l_1)} - d_{b_1}^{(l_1)} \right).
    \end{split}
    \label{eq:abbreviations}
\end{equation}

Clearly, the log-likelihood function in Eq.~(\ref{eq:loglikelihood}) is suited for Siamese neural network-based Channel Charting, where a loss is computed based on pairs of position estimates $\hat{\mathbf x}^{(l_1)}$ and $\hat{\mathbf x}^{(l_2)}$.
We remark that the phases $\varphi_b^{(l)}$, $b \in \{ 1, \ldots, B \}$, which were integrated and unwrapped along the \ac{UE}'s trajectory, may be interpreted as a \emph{path signature}.
In that sense, our approach shares some similarities with the concept of using path signatures for Channel Charting outlined in \cite{zhao2024signature}.

\subsection{Uncertainty}
Estimating the uncertainty in the phase difference measurement $\Delta \varphi_{b_1, b_2}^{(l_1, l_2)}$, described by $\sigma_{b_1, b_2}^{(l_1, l_2)}$, plays an important role in the definition of a suitable loss function $\mathcal L^{(l_1, l_2)}$.
For example, we must take into account that $\Delta \varphi_{b_1, b_2}^{(l_1, l_2)}$ is less precise if there is a long time delay $|t^{(l_2)} - t^{(l_1)}|$ between measurements (e.g., due to frequency drift, integration errors) and that the previously derived model for the relationship between displacement and phase only applies if the propagation channel is sufficiently \ac{LoS}-like.
Switching to continuous-time $t_1 = t^{(l_1)}$, $t_2 = t^{(l_2)}$ for notational reasons, we propose to compute $\sigma_{b_1, b_2}^{(l_1, l_2)}$ by using a heuristic that derives an instantaneous uncertainty $u_b(t) > 0$ from the delay spread of the channel between \ac{UE} and \ac{BS} antenna $b$.
That is, $u_b(t)$ is large for high delay spreads (non-\ac{LoS}) and small for low delay spreads (\ac{LoS}).
We then determine the standard deviation used for the loss function
\begin{equation}
    \begin{split}
        \sigma_{b_1, b_2}^{(l_1, l_2)} &= \beta + \int_{t_1}^{t_2} (u_{b_1}(t) + u_{b_2}(t)) ~ \mathrm dt \\
        &= \beta + \left(U_{b_1}(t_2) + U_{b_2}(t_2)\right) - \left(U_{b_1}(t_1) + U_{b_2}(t_1)\right)
    \end{split}\raisetag{1.8\baselineskip}
    \label{eq:uncertainty_int}
\end{equation}
by integrating over these instantaneous uncertainties, where $U_b(t)$ denotes the antiderivative of $u_b(t)$ and $\beta$ is a hyperparameter.
In practice, channel measurements and hence uncertainty heuristics $u_b^{(l)}$ are only available at discrete times $l$, hence Eq.~(\ref{eq:uncertainty_int}) turns into a summation and $U_b^{(l)}$ is found by numerical integration.
By first computing $U_b^{(l)}$ for all $b \in \{ 1, \ldots, B \}$, Eq.~(\ref{eq:uncertainty_int}) can be solved for arbitrary combinations of $b_1, b_2, l_1, l_2$ with low computational complexity.
The hyperparameter $\beta$ is chosen to be some small value, and can be interpreted as the minimum uncertainty about $\Delta \varphi_{b_1, b_2}^{(l_1, l_2)}$ even for small time differences.
It ensures that $\sigma_{b_1, b_2}^{(l_1, l_2)} > 0$ (no division by zero) and may be tweaked over training epochs.
Future work may improve this uncertainty estimation step.

\section{Foward Charting Function Training}
\label{sec:training}
We implement the \ac{FCF} $\mathcal C_{\theta, \mathrm{Dop}}: \mathbb C^{B \times N_\mathrm{sub}} \to \mathbb R^2$, which maps \ac{CSI} matrices $\mathbf H^{(l)}$ to positions $\hat{\mathbf x}^{(l)} = \mathcal C_{\theta, \mathrm{Dop}}\left(\mathbf H^{(l)}\right)$, as a \ac{NN}.
To use $\mathbf H^{(l)} \in \mathbb C^{B \times N_\mathrm{sub}}$ in a \ac{NN} that only operates on real numbers, an input layer extracts real part, imaginary part and logarithmic absolute values from a subset of time-domain channel taps $\mathcal F^{-1} \left\{\mathbf H^{(l)} \right\}$ (\acp{CIR}) for further processing.
The \ac{NN} is made up of 5 dense layers with \ac{ReLU} activation and 1024, 512, 256, 128 and 64 neurons each, followed by one output layer with linear activation and two neurons (two dimensions of position estimate $\hat {\mathbf x}$).
We train the Siamese neural network in six epochs with increasing batch sizes and decreasing learning rates.
In each epoch, the network is trained on $30\,000$ randomly selected pairs of datapoints.

Compared to previous work, e.g., \cite{fraunhofer_cc, stephan2023angle}, there is nothing special about the \ac{NN} architecture or training procedure.
However, in contrast to previous work, we use the negated log-likelihood function from Eq.~(\ref{eq:loglikelihood}) as a loss function for self-supervised training of the \ac{NN} in a Siamese \ac{NN} configuration.
Eq.~(\ref{eq:loglikelihood}) cannot be interpreted as a \emph{dissimilarity metric} (compare \cite{stephan2023angle}), i.e., as a spatial distance between datapoints $(l_1, l_2)$.
This is because $\mathcal L^{(l_1, l_2)}$ not just depends on the distance between the two datapoints, but on the absolute coordinates $\hat{\mathbf x}^{(l_1)}$ and $\hat{\mathbf x}^{(l_2)}$.
While this makes it harder to combine the Doppler effect-based Channel Charting approach with other approaches based on dissimilarity metrics, it enables the \ac{FCF} to learn not just relative locations, but to locate datapoints in the global coordinate frame.
Since Eq.~(\ref{eq:loglikelihood}) only operates on physically tangible quantities like distances and phase shifts, no rotation, scaling or translation (affine transformation) is required between channel chart and global coordinate frame.

\section{Evaluation}
\label{sec:evaluation}
\begin{table*}
    \centering
    \caption{Performance comparison: Doppler-based and dissimilarity metric-based Channel Charting}
    \vspace{-0.2cm}
    \begin{tabular}{r | c | c c c c c c c | l}
        & \textbf{Dataset} & \textbf{MAE $\downarrow$} & \textbf{DRMS $\downarrow$} & \textbf{CEP $\downarrow$} & \textbf{R95 $\downarrow$} & \textbf{CT $\uparrow$} & \textbf{TW $\uparrow$} & \textbf{KS $\downarrow$} & \textbf{Figure} \\ \hline
        $T_\mathrm{opt} \circ C_{\theta, \mathrm{Dop}}$ & $\mathcal S_\mathrm{train}$ & $1.033\,\mathrm m$ & $1.208\,\mathrm m$ & $0.912\,\mathrm m$ & $2.238\,\mathrm m$ & 0.955 & 0.953 & 0.165 & Fig.~\ref{fig:cc_tc_dop} \\
        $C_{\theta, \mathrm{Dop}}$ & $\mathcal S_\mathrm{train}$ & $\mathbf{1.235\,m}$ & $\mathbf{1.431\,m}$ & $\mathbf{1.108\,m}$ & $\mathbf{2.607\,m}$ & \textbf{0.953} & \textbf{0.951} & \textbf{0.170} & Fig.~\ref{fig:cc_dop} \\
        $T_\mathrm{opt} \circ C_{\theta, \mathrm{fuse}}$ & $\mathcal S_\mathrm{train}$ & $1.653\,\mathrm m$ & $1.923\,\mathrm m$ & $1.495\,\mathrm m$ & $3.541\,\mathrm m$ & 0.913 & 0.893 & 0.264 & Fig.~\ref{fig:cc_tc_fuse} \\
        $T_\mathrm{opt} \circ C_{\theta, \mathrm{CIRA}}$ & $\mathcal S_\mathrm{train}$ & $2.103\,\mathrm m$ & $2.345\,\mathrm m$ & $2.014\,\mathrm m$ & $4.112\,\mathrm m$ & 0.888 & 0.864 & 0.308 & Fig.~\ref{fig:cc_tc_cira} \\ \hline
        $C_{\theta, \mathrm{Dop}}$ & $\mathcal S_\mathrm{test}$ & $1.346\,\mathrm m$ & $1.540\,\mathrm m$ & $1.227\,\mathrm m$ & $2.757\,\mathrm m$ & 0.933 & 0.929 & 0.200 & --- \\
    \end{tabular}
    \label{tab:performance}
    \vspace{0.1cm}

    \textbf{MAE} = Mean Absolute Error, \textbf{DRMS} = Distance Root Mean Square, \textbf{CEP} = Circular Error Probable, \textbf{R95} = 95\textsuperscript{th} percentile radius,\newline
    \textbf{CT} = Continuity, \textbf{TW} = Trustworthiness, \textbf{KS} = Kruskal's Stress, all metrics as defined in \cite{stephan2023angle, asilomar2023}
    \vspace{-0.5cm}
\end{table*}

\begin{figure}
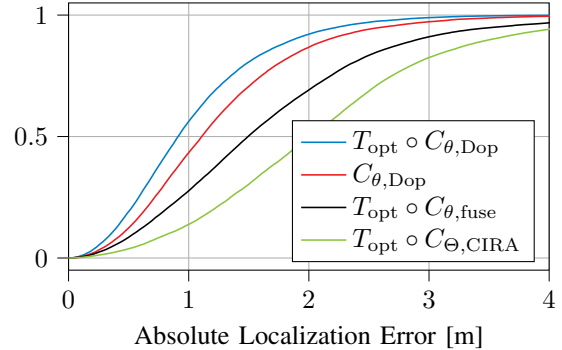

    \centering
    \include{fig/cdf}
    \vspace{-1.2cm}
    \caption{Empirical CDFs of absolute localization errors (evaluated on $\mathcal S_\mathrm{train}$)}
    \label{fig:cdf}
    \vspace{-0.3cm}
\end{figure}

\subsection{Baselines and Optimal Affine Transforms}
Finding baselines in existing Channel Charting literature to benchmark the proposed method against is challenging, since other state-of-the-art Channel Charting techniques assume time or phase synchronization and do not provide channel charts in the global coordinate frame.
In the absence of a better alternative, we use dissimilarity metric-based Channel Charting with the \ac{CIR} amplitude (CIRA) dissimilarity metric, proposed in \cite{fraunhofer_cc}, as a baseline.
Lacking time synchronization, we time-shift the \acp{CIR} for each \ac{BS} antenna individually such that the mean delay of the signal is zero.
Furthermore, as proposed in \cite{stephan2023angle}, we introduce a baseline where we fuse the CIRA metric with a timestamp-based metric.
In both cases, we use geodesic dissimilarities as in \cite{fraunhofer_cc}.
The trained \ac{NN}-based \acp{FCF} are called $C_{\theta, \mathrm{CIRA}}$ and $C_{\theta, \mathrm{fused}}$, respectively.
Since $C_{\theta, \mathrm{CIRA}}$ and $C_{\theta, \mathrm{fused}}$ produce channel charts whose coordinate frame may be arbitrarily transformed compared to the global coordinate frame, we find a linear mapping $T_\mathrm{opt}(\mathbf x) = \hat { \mathbf A } \mathbf x + \hat { \mathbf b }$ that describes the optimal affine transformation from channel chart coordinates to global coordinates by solving
\[
    (\mathbf{\hat A}, \mathbf{\hat b}) = \argmin\limits_{(\mathbf{A}, \mathbf{b})} \sum_{l = 1}^L \lVert\mathbf{A} \hat {\mathbf x }^{(l)} + \mathbf b - \mathbf x^{(l)} \rVert_2^2,
\]
where $\hat {\mathbf x }^{(l)} = \mathcal C_\theta\left(\mathbf H^{(l)}\right)$ denotes the position estimate obtained through Channel Charting, and $\mathbf x^{(l)}$ is the ``ground truth'' position from the dataset.
We write the composition of $T_\mathrm{opt}$ and $C_\theta$, i.e., application of $T_\mathrm{opt}$ after the \ac{FCF}, as $T_\mathrm{opt} \circ C_\theta$.
The performance for the baselines are computed \emph{after} applying $T_\mathrm{opt}$, which gives an unfair advantage to the baselines, but our approach outperforms the baselines regardless.

\subsection{Results}
Fig.~\ref{fig:cc-all} shows the learned channel charts, evaluated by applying the respective \ac{FCF} on $\mathcal S_\mathrm{train}$ and possibly also applying the respective optimal affine transform $T_\mathrm{opt}$.
Clearly, the proposed Doppler effect-based method manages to reconstruct the global geometry (L-shaped area) of the environment and even manages to locate datapoints in the global coordinate frame without subsequent transformation (Fig.~\ref{fig:cc_dop}).
As evidenced by Tab. \ref{tab:performance}, $C_{\theta, \mathrm{Dop}}$ not only produces visually more appealing channel charts, but also outperforms the baselines with respect to all considered performance metrics, even without applying $T_\mathrm{opt}$ (metrics highlighted in bold).
We want to stress that evaluating on the training set (and not only on a separate test set) is meaningful, since Channel Charting is a self-supervised learning technique, but the last row in Tab. \ref{tab:performance} indicates that performance on the test set is comparable anyway.
Finally, we consider the empirical distribution of the localization error $\left\lVert \mathbf x^{(l)} - \hat {\mathbf x}^{(l)} \right\rVert_2$, as shown in the \ac{eCDF} diagram in Fig.~\ref{fig:cdf}.
Unsurprisingly, $C_{\theta, \mathrm{Dop}}$ also outperforms the baselines in the \ac{eCDF}.


\vspace{-0.05cm}

\section{Outlook}
\label{sec:outlook}
Our results show that Channel Charting in global coordinates with only few \ac{BS} antennas is possible by exploiting the Doppler effect, a feature of wireless \ac{CSI} that has previously received little attention in the context of Channel Charting.
While we only use Doppler effect-induced frequency shifts to derive a loss function, taking the Doppler effect into account for Channel Charting opens up several other exciting research directions.
For example, one could derive a dissimilarity metric from frequency shifts, or use delay-Doppler domain features as \ac{FCF} inputs.
How to combine our Doppler effect-based loss function with dissimilarity-based or timestamp-based Channel Charting techniques remains an open question.

\vspace{-0.05cm}

\bibliographystyle{IEEEtran}
\bibliography{IEEEabrv,references}

\end{document}

%% file: acronyms.tex
\begin{acronym}
 \acro{R2M}{raw 2\textsuperscript{nd} moment}
 \acro{CSI}{channel state information}
 \acro{UE}{user equipment}
 \acro{EM}{electromagnetic}
 \acro{UL}{uplink}
 \acro{BS}{base station}
 \acro{TDD}{time division duplex}
 \acro{FDD}{frequency division duplex}
 \acro{ECC}{error-correcting code}
 \acro{MLD}{maximum likelihood decoding}
 \acro{HDD}{hard decision decoding}
 \acro{IF}{intermediate frequency}
 \acro{RF}{radio frequency}
 \acro{SDD}{soft decision decoding}
 \acro{NND}{neural network decoding}
 \acro{CNN}{convolutional neural network}
 \acro{ML}{maximum likelihood}
 \acro{GPU}{graphical processing unit}
 \acro{BP}{belief propagation}
 \acro{LTE}{Long Term Evolution}
 \acro{BER}{bit error rate}
 \acro{SNR}{signal-to-noise-ratio}
 \acro{ReLU}{rectified linear unit}
 \acro{BPSK}{binary phase shift keying}
 \acro{QPSK}{quadrature phase shift keying}
 \acro{AWGN}{additive white Gaussian noise}
 \acro{MSE}{mean squared error}
 \acro{LLR}{log-likelihood ratio}
 \acro{MAP}{maximum a posteriori}
 \acro{NVE}{normalized validation error}
 \acro{BCE}{binary cross-entropy}
 \acro{CE}{cross-entropy}
 \acro{BLER}{block error rate}
 \acro{SQR}{signal-to-quantisation-noise-ratio}
 \acro{MIMO}{multiple-input multiple-output}
 \acro{mMIMO}{massive multiple-input multiple-output}
 \acro{OFDM}{orthogonal frequency division multiplex}
 \acro{RF}{radio frequency}
 \acro{LoS}{line of sight}
 \acro{NLoS}{non-line of sight}
 \acro{NMSE}{normalized mean squared error}
 \acro{CFO}{carrier frequency offset}
 \acro{SFO}{sampling frequency offset}
 \acro{IPS}{indoor positioning system}
 \acro{TRIPS}{time-reversal IPS}
 \acro{RSSI}{received signal strength indicator}
 \acro{MIMO}{multiple-input multiple-output}
 \acro{ENoB}{effective number of bits}
 \acro{AGC}{automatic gain control}
 \acro{ADC}{analog to digital converter}
 \acro{ADCs}{analog to digital converters}
 \acro{FB}{front bandpass}
 \acro{FPGA}{field programmable gate array}
 \acro{JSDM}{Joint Spatial Division and Multiplexing}
 \acro{NN}{neural network}
 \acro{IF}{intermediate frequency}
 \acro{LoS}{line-of-sight}
 \acro{NLoS}{non-line-of-sight}
 \acro{DSP}{digital signal processing}
 \acro{AFE}{analog front end}
 \acro{SQNR}{signal-to-quantisation-noise-ratio}
 \acro{SINR}{signal-to-interference-noise-ratio}
 \acro{ENoB}{effective number of bits}
 \acro{PCB}{printed circuit board}
 \acro{EVM}{error vector mangnitude}
 \acro{CDF}{cumulative distribution function}
 \acro{MRC}{maximum ratio combining}
 \acro{MRP}{maximum ratio precoding}
 \acro{MRT}{maximum ratio transmission}
 \acro{DeepL}{deep-learning}
 \acro{DL}{downlink}
 \acro{SISO}{single-input single-output}
 \acro{SGD}{stochastic gradient descent}
 \acro{CP}{cyclic prefix}
 \acro{MISO}{Multiple Input Single Output}
 \acro{LMMSE}{linear minimum mean square error}
 \acro{ZF}{zero forcing}
 \acro{USRP}{universal software radio peripheral}
 \acro{RNN}{recurrent neural network}
 \acro{GRU}{gated recurrent unit}
 \acro{LSTM}{long short-term memory}
 \acro{NTM}{neural turing machine}
 \acro{DNC}{differentiable neural computer}
 \acro{TCN}{temporal convolutional network}
 \acro{FCL}{fully connected layer}
 \acro{MANN}{memory augmented neural network}
 \acro{RNN}{recurrent neural network}
 \acro{DNN}{deep neural network}
 \acro{FIR}{finite impulse response}
 \acro{BPTT}{back-propagation through time}
 \acro{GAN}{generative adversarial network}
 \acro{ELU}{exponential linear unit}
 \acro{tanh}{hyperbolic tangent}
 \acro{BICM}{bit-interleaved coded modulation}
 \acro{OTA}{over-the-air}
 \acro{IM}{intensity modulation}
 \acro{DD}{direct detection}
 \acro{RL}{reinforcement learning}
 \acro{SDR}{software-defined radio}
 \acro{WGAN}{Wasserstein generative adversarial network}
 \acro{BMD}{bit-metric decoding}
 \acro{BMI}{bit-wise mutual information}
 \acro{LDPC}{low-density parity-check}
 \acro{IDD}{iterative demapping and decoding}
 \acro{JSD}{Jensen-Shannon divergence}
 \acro{MMSE}{minimum mean square error}
 \acro{FFT}{fast Fourier transform}
 \acro{DFT}{discrete Fourier transform}
 \acro{IFFT}{inverse fast Fourier transform}
 \acro{QAM}{quadrature amplitude modulation}
 \acro{EMD}{earth mover's distance}
 \acro{TDL}{tapped delay line}
 \acro{KL}{Kullback-Leibler}
 \acro{PRACH}{physical random access channel}
 \acro{URLLC}{ultra-reliable low-latency communication}
 \acro{ANOMA}{asynchronous non-orthogonal multiple access}
 \acro{FEC}{forward error correction}
 \acro{PAPR}{peak-to-average power ratio}
 \acro{APP}{a posteriori probability}
 \acro{COTS}{commercial off-the-shelf}
 \acro{PLL}{phase locked loop}
 \acro{STO}{sampling time offset}
 \acro{SFO}{sampling frequency offset}
 \acro{CFO}{carrier frequency offset}
 \acro{CPO}{carrier phase offset}
 \acro{CSI}{channel state information}
 \acro{GNSS}{global navigation satellite system}
 \acro{ELAA}{extremely large aperture array}
 \acro{UE}{user equipment}
 \acro{DICHASUS}{\underline{Di}stributed \underline{Cha}nnel \underline{S}ounder by \underline{U}niversity of \underline{S}tuttgart}
 \acro{JCaS}{Joint Communication and Sensing}
 \acro{CT}{continuity}
 \acro{TW}{trustworthiness}
 \acro{KS}{Kruskal's stress}
 \acro{PCA}{principal component analysis}
 \acro{AoA}{angle of arrival}
 \acro{ToA}{time of arrival}
 \acro{TDoA}{time difference of arrival}
 \acro{ToT}{time of transmission}
 \acro{RD}{Rajski's distance}
 \acro{ADP}{angle-delay profile}
 \acro{MPC}{multipath component}
 \acro{CIR}{channel impulse response}
 \acro{MAE}{mean absolute error}
 \acro{MDS}{multidimensional scaling}
 \acro{t-SNE}{t-distributed stochastic neighbor embedding}
 \acro{SM}{Sammon's mapping}
 \acro{CIRA}{channel impulse response amplitude}
 \acro{CS}{cosine similarity}
 \acro{FCF}{forward charting function}
 \acro{CMD}{correlation matrix distance}
 \acro{UWB}{ultra-wideband}
 \acro{MUSIC}{MUltiple SIgnal Classification}
 \acro{FBCM}{forward-backward correlation matrix}
 \acro{MDL}{minimum descriptive length}
 \acro{BFGS}{Broyden-Fletcher-Goldfarb-Shanno}
 \acro{UPA}{uniform planar array}
 \acro{RMS}{root mean square}
 \acro{DRMS}{distance root mean square}
 \acro{CEP}{circular error probable}
 \acro{R95}{95\textsuperscript{th} percentile radius}
 \acro{eCDF}{empirical cumulative distribution function}
 \acro{GPSDO}{GPS disciplined oscillator}
\end{acronym}

%% file: corporate_colors.tex
\definecolor{mittelblau}{RGB}{0, 126, 198}
\definecolor{violettblau}{cmyk}{0.9, 0.6, 0, 0}
\definecolor{rot}{RGB}{238, 28 35}
\definecolor{apfelgruen}{RGB}{140, 198, 62}
\definecolor{gelb}{RGB}{1, 221, 0}
\definecolor{orange}{RGB}{244, 111, 33}
\definecolor{pink}{RGB}{237, 0, 140}
\definecolor{lila}{RGB}{128, 10, 145}
\definecolor{hellgrau}{RGB}{224, 224, 224}
\definecolor{mittelgrau}{RGB}{128, 128, 128}
\definecolor{dunkelgrau}{RGB}{80,80,80}
\definecolor{anthrazit}{RGB}{19, 31, 31}

%% file: fig/cdf.tex
\newif\ifsupervised
\newif\ifccaugmented
\newif\ifcccheat
\newif\ifccclassical
\newif\ifelevation
\newif\ifjoint
\newif\ifaoa
\newif\iftoa
\newif\iftaner

\supervisedfalse
\ccaugmentedtrue
\cccheatfalse
\ccclassicaltrue
\jointtrue
\aoatrue
\elevationfalse
\toatrue
\tanerfalse

\begin{tikzpicture}

\definecolor{darkgray176}{RGB}{176,176,176}
\definecolor{steelblue31119180}{RGB}{31,119,180}

\definecolor{mittelblau}{RGB}{0, 126, 198}
\definecolor{violettblau}{cmyk}{0.9, 0.6, 0, 0}
\definecolor{rot}{RGB}{238, 28 35}
\definecolor{apfelgruen}{RGB}{140, 198, 62}
\definecolor{gelb}{RGB}{1, 221, 0}
\definecolor{orange}{RGB}{244, 111, 33}
\definecolor{pink}{RGB}{237, 0, 140}
\definecolor{lila}{RGB}{128, 10, 145}
\definecolor{hellgrau}{RGB}{224, 224, 224}
\definecolor{mittelgrau}{RGB}{128, 128, 128}
\definecolor{dunkelgrau}{RGB}{80,80,80}
\definecolor{anthrazit}{RGB}{19, 31, 31}

\pgfplotsset{
	cycle list/Dark2-6,
	cycle multiindex* list={
		mark list*\nextlist
		Dark2-6\nextlist
	},
}

\begin{axis}[
tick align=outside,
width=0.9\columnwidth,height=0.58\columnwidth,
tick pos=left,
x grid style={darkgray176},
xlabel={Absolute Localization Error [m]},
xmajorgrids,
xmin=0, xmax=4,
xtick style={color=black},
y grid style={darkgray176},
ymajorgrids,
ymin=-0.05, ymax=1.05,
ytick style={color=black},
legend pos = south east,
legend cell align={left}
]

\addplot [semithick, mittelblau]
table {%
0 0
0.0291164776001611 0.000767349287803942
0.0546848434871936 0.00273368183780154
0.0802532093742261 0.00546736367560309
0.105821575261259 0.00973574408901252
0.131389941148291 0.0142439211548607
0.156958307035324 0.0209582274231452
0.182526672922356 0.0271929403865522
0.208095038809389 0.0362572538487363
0.233663404696421 0.0452736079804326
0.259231770583454 0.0565920099755407
0.284800136470486 0.066375713395041
0.310368502357519 0.0781257493645389
0.335936868244551 0.0900676226559877
0.361505234131584 0.103879909836459
0.387073600018616 0.117884034338881
0.412641965905649 0.133087142103496
0.438210331792681 0.150496379070548
0.463778697679714 0.167138266749796
0.489347063566746 0.184787300369287
0.514915429453779 0.199798570811951
0.540483795340811 0.217831279075344
0.566052161227844 0.2360078653302
0.591620527114876 0.25600690614359
0.617188893001909 0.276629418253321
0.642757258888941 0.296052947100858
0.668325624775974 0.314709126660592
0.693893990663006 0.33518776077886
0.719462356550039 0.35581027288859
0.745030722437071 0.375425639058079
0.770599088324104 0.394849167905616
0.796167454211137 0.414224737422666
0.821735820098169 0.433360510287276
0.847304185985202 0.452400364490912
0.872872551872234 0.470289194762841
0.898440917759267 0.490815788211597
0.924009283646299 0.508992374466452
0.949577649533332 0.528511821974965
0.975146015420364 0.545681262289578
1.0007143813074 0.562658865282241
1.02628274719443 0.578869118987099
1.05185111308146 0.594359982734641
1.07741947896849 0.609754927821208
1.10298784485553 0.625533547551676
1.12855621074256 0.640305021341902
1.15412457662959 0.655987722411395
1.17969294251662 0.670279602896744
1.20526130840366 0.682557191501607
1.23082967429069 0.696033763368663
1.25639804017772 0.707112368711333
1.28196640606475 0.719917509951561
1.30753477195179 0.73080427797228
1.33310313783882 0.74164308666251
1.35867150372585 0.753441081962496
1.38423986961288 0.763128866721021
1.40980823549992 0.77463910603808
1.43537660138695 0.784470768788068
1.46094496727398 0.793870797563666
1.48651333316101 0.802695314373411
1.51208169904805 0.810512685242914
1.53765006493508 0.819337202052659
1.56321843082211 0.826866816939236
1.58878679670914 0.834588269147763
1.61435516259618 0.841062778763609
1.63992352848321 0.84763320704043
1.66549189437024 0.853963838664812
1.69106026025727 0.86082202292456
1.71662862614431 0.866960817226992
1.74219699203134 0.872476140233083
1.76776535791837 0.877511869934296
1.7933337238054 0.882787396287948
1.81890208969244 0.888542515946478
1.84447045557947 0.894009879622081
1.8700388214665 0.898805812670855
1.89560718735353 0.903937461033044
1.92117555324057 0.908397678768405
1.9467439191276 0.912905855834253
1.97231228501463 0.917126276917174
1.99788065090166 0.920915064025706
2.0234490167887 0.924991607117165
2.04901738267573 0.928636516234233
2.07458574856276 0.932617140664716
2.10015411444979 0.93530286317203
2.12572248033683 0.938468178984222
2.15129084622386 0.941105942161048
2.17685921211089 0.943983501990312
2.20242757799792 0.946429427845187
2.22799594388496 0.948875353700062
2.25356430977199 0.95189679152079
2.27913267565902 0.954102920723227
2.30470104154605 0.956404968586639
2.33026940743309 0.958802935111026
2.35583777332012 0.960913145652487
2.38140613920715 0.962783559541509
2.40697450509418 0.964510095439068
2.43254287098122 0.966284590667114
2.45811123686825 0.967675411251259
2.48367960275528 0.969258069157355
2.50924796864231 0.971128483046377
2.53481633452935 0.972855018943936
2.56038470041638 0.974293798858568
2.58595306630341 0.975396863459786
2.61152143219044 0.97616421274759
2.63708979807748 0.976739724713443
2.66265816396451 0.977890748645149
2.68822652985154 0.978897894585392
2.71379489573857 0.979905040525634
2.73936326162561 0.980816267804901
2.76493162751264 0.982207088389046
2.79049999339967 0.983070356337826
2.8160683592867 0.984077502278068
2.84163672517374 0.985228526209774
2.86720509106077 0.985995875497578
2.8927734569478 0.986667306124407
2.91834182283483 0.987530574073186
2.94391018872187 0.988106086039039
2.9694785546089 0.988873435326843
2.99504692049593 0.989400987962208
3.02061528638296 0.989928540597573
3.04618365227 0.990264255910988
3.07175201815703 0.990887727207328
3.09732038404406 0.991319361181718
3.1228887499311 0.991894873147571
3.14845711581813 0.992326507121961
3.17402548170516 0.992854059757326
3.19959384759219 0.993381612392691
3.22516221347923 0.993525490384154
3.25073057936626 0.994101002350007
3.27629894525329 0.994484676993909
3.30186731114032 0.994868351637811
3.32743567702735 0.99510814829025
3.35300404291439 0.995299985612201
3.37857240880142 0.99553978226464
3.40414077468845 0.995923456908542
3.42970914057548 0.99616325356098
3.45527750646252 0.996355090882931
3.48084587234955 0.996451009543907
3.50641423823658 0.996690806196346
3.53198260412362 0.996930602848784
3.55755097001065 0.996978562179272
3.58311933589768 0.997122440170735
3.60868770178471 0.997362236823174
3.63425606767175 0.997506114814637
3.65982443355878 0.997745911467076
3.68539279944581 0.997841830128051
3.71096116533284 0.998033667450002
3.73652953121988 0.998129586110978
3.76209789710691 0.998129586110978
3.78766626299394 0.998225504771953
3.81323462888097 0.998321423432929
3.83880299476801 0.998417342093904
3.86437136065504 0.998465301424392
3.88993972654207 0.99851326075488
3.9155080924291 0.998609179415855
3.94107645831614 0.998657138746343
3.96664482420317 0.998753057407318
3.9922131900902 0.998801016737806
4.01778155597723 0.998801016737806
4.04334992186427 0.998896935398782
4.0689182877513 0.998992854059757
4.09448665363833 0.999088772720733
4.12005501952536 0.999088772720733
4.1456233854124 0.99913673205122
4.17119175129943 0.99913673205122
4.19676011718646 0.999184691381708
4.22232848307349 0.999232650712196
4.24789684896053 0.999232650712196
4.27346521484756 0.999232650712196
4.29903358073459 0.999280610042684
4.32460194662162 0.999424488034147
4.35017031250866 0.999424488034147
4.37573867839569 0.999472447364634
4.40130704428272 0.999472447364634
4.42687541016975 0.999520406695122
4.45244377605679 0.99956836602561
4.47801214194382 0.999664284686585
4.50358050783085 0.999664284686585
4.52914887371788 0.999664284686585
4.55471723960492 0.999664284686585
4.58028560549195 0.999760203347561
4.60585397137898 0.999760203347561
4.63142233726601 0.999808162678049
4.65699070315305 0.999856122008536
4.68255906904008 0.999856122008536
4.70812743492711 0.999856122008536
4.73369580081414 0.999856122008536
4.75926416670118 0.999856122008536
4.78483253258821 0.999856122008536
4.81040089847524 0.999856122008536
4.83596926436227 0.999856122008536
4.86153763024931 0.999856122008536
4.88710599613634 0.999904081339024
4.91267436202337 0.999904081339024
4.9382427279104 0.999904081339024
4.96381109379744 0.999904081339024
4.98937945968447 0.999904081339024
5.0149478255715 0.999904081339024
5.04051619145853 0.999952040669512
5.06608455734557 0.999952040669512
5.0916529232326 0.999952040669512
5.11722128911963 1
};
\addlegendentry{$T_\mathrm{opt} \circ C_{\theta, \mathrm{Dop}}$}

\addplot [semithick, rot]
table {%
0 0
0.0319058309935691 0.00052755263536521
0.0595119953024445 0.0019663325499976
0.0871181596113199 0.00388470576950746
0.114724323920195 0.00685818425974773
0.142330488229071 0.0103592153853532
0.169936652537946 0.0149153517816891
0.197542816846821 0.0198072034914393
0.225148981155697 0.0254184451585056
0.252755145464572 0.033283775358496
0.280361309773448 0.0406695122536089
0.307967474082323 0.0480552491487219
0.335573638391199 0.056927725288955
0.363179802700074 0.0664236727255288
0.390785967008949 0.0771665627547839
0.418392131317825 0.0872380221572107
0.4459982956267 0.0987003021437821
0.473604459935576 0.111745240036449
0.501210624244451 0.124214665963263
0.528816788553326 0.138218790465685
0.556422952862202 0.15236679295957
0.584029117171077 0.167857656707112
0.611635281479953 0.18387607309002
0.639241445788828 0.200230204786341
0.666847610097703 0.216968011126565
0.694453774406579 0.23495276005947
0.722059938715454 0.253513020958227
0.74966610302433 0.270250827298451
0.777272267333205 0.287947820248429
0.80487843164208 0.307419308426454
0.832484595950956 0.326123447316675
0.860090760259831 0.344683708215433
0.887696924568707 0.361613351877608
0.915303088877582 0.379550141480025
0.942909253186457 0.397726727734881
0.970515417495333 0.41561555800681
0.998121581804208 0.432593160999472
1.02572774611308 0.449187089348233
1.05333391042196 0.465445302383579
1.08094007473083 0.481703515418925
1.10854623903971 0.500071938995732
1.13615240334859 0.517625053954247
1.16375856765746 0.534506738285933
1.19136473196634 0.551484341278596
1.21897089627521 0.567454798331015
1.24657706058409 0.582849743417582
1.27418322489296 0.59800489185171
1.30178938920184 0.612968202963887
1.32939555351071 0.626684571483382
1.35700171781959 0.640688695985804
1.38460788212846 0.654357105174812
1.41221404643734 0.668457148338209
1.43982021074621 0.68121433024795
1.46742637505509 0.692772528895496
1.49503253936397 0.704138890221092
1.52263870367284 0.716224641504004
1.55024486798172 0.728118555464966
1.57785103229059 0.739197160807635
1.60545719659947 0.750611481463719
1.63306336090834 0.760347225552731
1.66066952521722 0.771761546208815
1.68827568952609 0.781065656323438
1.71588185383497 0.790705481751475
1.74348801814384 0.799242242578293
1.77109418245272 0.807251450769747
1.79870034676159 0.815980048918517
1.82630651107047 0.823461704474605
1.85391267537935 0.831566831327035
1.88151883968822 0.838856649561172
1.9091250039971 0.845858711812383
1.93673116830597 0.85358016402091
1.96433733261485 0.860150592297731
1.99194349692372 0.865665915303822
2.0195496612326 0.872476140233082
2.04715582554147 0.878998609179416
2.07476198985035 0.884130257541605
2.10236815415922 0.889213946573306
2.1299743184681 0.893866001630617
2.15758048277698 0.898997649992806
2.18518664708585 0.903409908397679
2.21279281139473 0.907438492158649
2.2403989757036 0.911994628554985
2.26800514001248 0.915687497002542
2.29561130432135 0.919236487458635
2.32321746863023 0.922305884609851
2.3508236329391 0.92551915975253
2.37842979724798 0.929739580835451
2.40603596155685 0.932281425351302
2.43364212586573 0.93573449714642
2.4612482901746 0.938372260323246
2.48885445448348 0.940866145508609
2.51646061879236 0.943264112032996
2.54406678310123 0.945566159896408
2.57167294741011 0.94743657378543
2.59927911171898 0.949498824996403
2.62688527602786 0.951177401563474
2.65449144033673 0.953047815452496
2.68209760464561 0.954774351350055
2.70970376895448 0.9561651719342
2.73730993326336 0.958131504484197
2.76491609757223 0.960097837034195
2.79252226188111 0.961728454270778
2.82012842618999 0.963550908829313
2.84773459049886 0.964749892091506
2.87534075480774 0.966572346650041
2.90294691911661 0.968107045225648
2.93055308342549 0.969401947148818
2.95815924773436 0.970552971080524
2.98576541204324 0.972039710325644
3.01337157635211 0.972950937604911
3.04097774066099 0.973958083545154
3.06858390496986 0.974917270154908
3.09619006927874 0.97616421274759
3.12379623358761 0.976931562035394
3.15140239789649 0.978274423289051
3.17900856220536 0.979857081195146
3.20661472651424 0.981151983118316
3.23422089082312 0.982207088389046
3.26182705513199 0.983022397007338
3.28943321944087 0.983358112320752
3.31703938374974 0.984077502278068
3.34464554805862 0.984844851565872
3.37225171236749 0.985420363531725
3.39985787667637 0.985851997506115
3.42746404098524 0.986427509471968
3.45507020529412 0.986955062107333
3.48267636960299 0.987482614742698
3.51028253391187 0.988441801352453
3.53788869822075 0.988921394657331
3.56549486252962 0.989592825284159
3.5931010268385 0.989736703275622
3.62070719114737 0.990264255910987
3.64831335545625 0.990791808546353
3.67591951976512 0.991175483190255
3.703525684074 0.991750995156107
3.73113184838287 0.992134669800009
3.75873801269175 0.99232650712196
3.78634417700062 0.992806100426838
3.8139503413095 0.993237734401228
3.84155650561837 0.993525490384154
3.86916266992725 0.994005083689032
3.89676883423613 0.994388758332933
3.924374998545 0.994532636324397
3.95198116285388 0.994772432976835
3.97958732716275 0.99510814829025
4.00719349147163 0.995299985612201
4.0347996557805 0.99573161958659
4.06240582008938 0.996019375569517
4.09001198439825 0.996355090882931
4.11761814870713 0.996690806196345
4.145224313016 0.996882643518296
4.17283047732488 0.996978562179272
4.20043664163375 0.996978562179272
4.22804280594263 0.997506114814637
4.25564897025151 0.997554074145125
4.28325513456038 0.997793870797564
4.31086129886926 0.997841830128051
4.33846746317813 0.997889789458539
4.36607362748701 0.998033667450002
4.39367979179588 0.99808162678049
4.42128595610476 0.998225504771953
4.44889212041363 0.998273464102441
4.47649828472251 0.998465301424392
4.50410444903138 0.99851326075488
4.53171061334026 0.998657138746343
4.55931677764914 0.998657138746343
4.58692294195801 0.998657138746343
4.61452910626689 0.99870509807683
4.64213527057576 0.998896935398781
4.66974143488464 0.998944894729269
4.69734759919351 0.998944894729269
4.72495376350239 0.998944894729269
4.75255992781126 0.999088772720732
4.78016609212014 0.999088772720732
4.80777225642901 0.99913673205122
4.83537842073789 0.999280610042683
4.86298458504676 0.999280610042683
4.89059074935564 0.999328569373171
4.91819691366452 0.999328569373171
4.94580307797339 0.999376528703659
4.97340924228227 0.999376528703659
5.00101540659114 0.999424488034147
5.02862157090002 0.999472447364634
5.05622773520889 0.99956836602561
5.08383389951777 0.999760203347561
5.11144006382664 0.999760203347561
5.13904622813552 0.999808162678048
5.16665239244439 0.999856122008536
5.19425855675327 0.999856122008536
5.22186472106215 0.999856122008536
5.24947088537102 0.999856122008536
5.2770770496799 0.999856122008536
5.30468321398877 0.999856122008536
5.33228937829765 0.999856122008536
5.35989554260652 0.999904081339024
5.3875017069154 0.999904081339024
5.41510787122427 0.999904081339024
5.44271403553315 0.999952040669512
5.47032019984202 0.999952040669512
5.4979263641509 0.999952040669512
5.52553252845977 0.999999999999999
};
\addlegendentry{$C_{\theta, \mathrm{Dop}}$}

\addplot [semithick, black]
table {%
0 0
0.0340631315679364 0.000575511965852957
0.065341030607928 0.00177449522804662
0.0966189296479196 0.00374082777804422
0.127896828687911 0.00628267229389478
0.159174727727903 0.0093041101146228
0.190452626767894 0.0134765718670567
0.221730525807886 0.0187041388902211
0.253008424847877 0.0244592585487507
0.284286323887869 0.0299745815548415
0.315564222927861 0.035921538535322
0.346842121967852 0.043499112752386
0.378120021007844 0.0513164836218886
0.409397920047835 0.0591338544913913
0.440675819087827 0.0673349000047959
0.471953718127818 0.0765430914584432
0.50323161716781 0.0864706728694067
0.534509516207801 0.0973094815596374
0.565787415247793 0.107428900292552
0.597065314287785 0.118843220948636
0.628343213327776 0.129250395664477
0.659621112367768 0.14066471632056
0.690899011407759 0.152318833629083
0.722176910447751 0.166131120809553
0.753454809487742 0.177257685482711
0.784732708527734 0.189727111409525
0.816010607567725 0.202628171310728
0.847288506607717 0.215001678576567
0.878566405647709 0.226703755215577
0.9098443046877 0.239508896455805
0.941122203727692 0.252649753009448
0.972400102767683 0.264975300944799
1.00367800180767 0.278068198167954
1.03495590084767 0.292120282000863
1.06623379988766 0.306747877799626
1.09751169892765 0.319696897031317
1.12878959796764 0.33432449283008
1.16006749700763 0.34688983741787
1.19134539604762 0.361325595894681
1.22262329508762 0.37561747638003
1.25390119412761 0.390245072178792
1.2851790931676 0.405448179943408
1.31645699220759 0.419979857081195
1.34773489124758 0.43412785957508
1.37901279028757 0.448275862068966
1.41029068932757 0.462327945901875
1.44156858836756 0.477003501031126
1.47284648740755 0.490432113567695
1.50412438644754 0.504723994053043
1.53540228548753 0.5182005659201
1.56668018452752 0.531149585151791
1.59795808356751 0.544530238357873
1.62923598260751 0.556615989640785
1.6605138816475 0.568509903601746
1.69179178068749 0.581027288859048
1.72306967972748 0.593448755455374
1.75434757876747 0.605966140712676
1.78562547780746 0.617476380029735
1.81690337684746 0.628554985372404
1.84818127588745 0.639873387367513
1.87945917492744 0.651575464006523
1.91073707396743 0.6618387607309
1.94201497300742 0.672629610090643
1.97329287204741 0.682940866145509
2.0045707710874 0.693300081530862
2.0358486701274 0.704186849551581
2.06712656916739 0.714594024267421
2.09840446820738 0.72399405304302
2.12968236724737 0.734065512445446
2.16096026628736 0.74356145988202
2.19223816532735 0.753345163301521
2.22351606436735 0.763608460025898
2.25479396340734 0.773248285453935
2.28607186244733 0.782360558246607
2.31734976148732 0.79123303438684
2.34862766052731 0.799577957891708
2.3799055595673 0.806292264159993
2.4111834586073 0.813630041724618
2.44246135764729 0.821591290585584
2.47373925668728 0.829024986811184
2.50501715572727 0.835547455757518
2.53629505476726 0.841782168720925
2.56757295380725 0.848112800345307
2.59885085284724 0.853532204690423
2.63012875188724 0.858999568366026
2.66140665092723 0.863843460745288
2.69268454996722 0.86911898709894
2.72396244900721 0.873531245503813
2.7552403480472 0.878375137883075
2.78651824708719 0.883075152270875
2.81779614612719 0.887439451345259
2.84907404516718 0.891324157114767
2.88035194420717 0.896024171502566
2.91162984324716 0.900148673924512
2.94290774228715 0.904129298354995
2.97418564132714 0.907630329480601
3.00546354036713 0.911275238597669
3.03674143940713 0.914344635748885
3.06801933844712 0.917749748213515
3.09929723748711 0.920723226703756
3.1305751365271 0.923744664524484
3.16185303556709 0.926094671718383
3.19313093460708 0.92935590619155
3.22440883364708 0.931609994724474
3.25568673268707 0.934295717231788
3.28696463172706 0.935974293798859
3.31824253076705 0.938180423001295
3.34952042980704 0.940434511534219
3.38079832884703 0.942017169440315
3.41207622788703 0.943791664668362
3.44335412692702 0.945422281904945
3.47463202596701 0.94681310248909
3.505909925007 0.948203923073234
3.53718782404699 0.949738621648842
3.56846572308698 0.951081482902499
3.59974362212697 0.952328425495181
3.63102152116697 0.953527408757374
3.66229942020696 0.954918229341519
3.69357731924695 0.956261090595176
3.72485521828694 0.958275382475661
3.75613311732693 0.959378447076879
3.78741101636692 0.960721308330536
3.81868891540692 0.961536616948828
3.84996681444691 0.962639681550046
3.8812447134869 0.963550908829313
3.91252261252689 0.964653973430531
3.94380051156688 0.965804997362237
3.97507841060687 0.967051939954919
4.00635630964686 0.968538679200039
4.03763420868686 0.969306028487843
4.06891210772685 0.970409093089061
4.10019000676684 0.971512157690279
4.13146790580683 0.97275910028296
4.16274580484682 0.973718286892715
4.19402370388681 0.974965229485397
4.22530160292681 0.975876456764664
4.2565795019668 0.976787684043931
4.28785740100679 0.977602992662223
4.31913530004678 0.978274423289051
4.35041319908677 0.979473406551245
4.38169109812676 0.980528511821975
4.41296899716676 0.981151983118316
4.44424689620675 0.981775454414657
4.47552479524674 0.982878519015875
4.50680269428673 0.983549949642703
4.53808059332672 0.984173420939044
4.56935849236671 0.984940770226848
4.6006363914067 0.985612200853676
4.6319142904467 0.986763224785382
4.66319218948669 0.987770370725625
4.69447008852668 0.988489760682941
4.72574798756667 0.989113231979282
4.75702588660666 0.989640784614647
4.78830378564665 0.990312215241475
4.81958168468665 0.990935686537816
4.85085958372664 0.991415279842693
4.88213748276663 0.991655076495132
4.91341538180662 0.99213466980001
4.94469328084661 0.992662222435375
4.9759711798866 0.993237734401228
5.00724907892659 0.993525490384154
5.03852697796659 0.993861205697569
5.06980487700658 0.994244880341471
5.10108277604657 0.994628554985373
5.13236067508656 0.995012229629274
5.16363857412655 0.995252026281713
5.19491647316654 0.995587741595128
5.22619437220654 0.995731619586591
5.25747227124653 0.99597141623903
5.28875017028652 0.99616325356098
5.32002806932651 0.996498968874395
5.3513059683665 0.996786724857321
5.38258386740649 0.997170399501223
5.41386176644649 0.997314277492686
5.44513966548648 0.997362236823174
5.47641756452647 0.997697952136588
5.50769546356646 0.997985708119515
5.53897336260645 0.998129586110978
5.57025126164644 0.998225504771954
5.60152916068643 0.998273464102441
5.63280705972643 0.998369382763417
5.66408495876642 0.998561220085368
5.69536285780641 0.998609179415855
5.7266407568464 0.998705098076831
5.75791865588639 0.998753057407319
5.78919655492638 0.998801016737806
5.82047445396638 0.99894489472927
5.85175235300637 0.999040813390245
5.88303025204636 0.999232650712196
5.91430815108635 0.999424488034147
5.94558605012634 0.99956836602561
5.97686394916633 0.999616325356098
6.00814184820632 0.999616325356098
6.03941974724632 0.999616325356098
6.07069764628631 0.999664284686586
6.1019755453263 0.999760203347561
6.13325344436629 0.999760203347561
6.16453134340628 0.999856122008537
6.19580924244627 0.999904081339024
6.22708714148627 0.999952040669512
6.25836504052626 1
};
\addlegendentry{$T_\mathrm{opt} \circ C_{\theta, \mathrm{fuse}}$}

\addplot [semithick, apfelgruen]
table {%
0 0
0.0452661324718769 0.000383674643901971
0.0743585545703563 0.00105510527073042
0.103450976668836 0.00211021054146084
0.132543398767315 0.00388470576950746
0.161635820865794 0.00575511965852957
0.190728242964274 0.00772145220852717
0.219820665062753 0.00935206944511054
0.248913087161232 0.0119898326219366
0.278005509259712 0.0142439211548607
0.307097931358191 0.0164020910268093
0.336190353456671 0.0189439355426598
0.36528277555515 0.0217255767109491
0.394375197653629 0.0253225264975301
0.423467619752109 0.0282960049877704
0.452560041850588 0.0316531581219126
0.481652463949067 0.0355858232219078
0.510744886047547 0.039710325643854
0.539837308146026 0.0433552347609227
0.568929730244506 0.0480552491487219
0.598022152342985 0.053330775502374
0.627114574441464 0.0574552779243202
0.656206996539944 0.0640257062011414
0.685299418638423 0.0703083784950362
0.714391840736902 0.0757277828401515
0.743484262835382 0.0822502517864851
0.772576684933861 0.0880533307755024
0.80166910703234 0.0939043690950074
0.83076152913082 0.100139082058414
0.859853951229299 0.105654405064505
0.888946373327779 0.111217687401084
0.918038795426258 0.118123830991319
0.947131217524737 0.124790177929116
0.976223639623217 0.132847345451058
1.0053160617217 0.13960961104983
1.03440848382018 0.147187185266894
1.06350090591865 0.155004556136396
1.09259332801713 0.164404584911995
1.12168575011561 0.173277061052228
1.15077817221409 0.182149537192461
1.17987059431257 0.191597525298547
1.20896301641105 0.200757757421706
1.23805543850953 0.211308810129011
1.26714786060801 0.22133231020095
1.29624028270649 0.230396623663134
1.32533270480497 0.240516042396048
1.35442512690345 0.252218119035058
1.38351754900193 0.263344683708215
1.41260997110041 0.273224305788691
1.44170239319889 0.2821927005899
1.47079481529737 0.292887631288667
1.49988723739585 0.303438683995971
1.52897965949432 0.315716272600834
1.5580720815928 0.326890796604479
1.58716450369128 0.339168385209342
1.61625692578976 0.350678624526402
1.64534934788824 0.361229677233706
1.67444176998672 0.372596038559302
1.7035341920852 0.383195050597094
1.73262661418368 0.392786916694643
1.76171903628216 0.403625725384874
1.79081145838064 0.41518392403242
1.81990388047912 0.426214570044602
1.8489963025776 0.438683995971416
1.87808872467608 0.450050357297012
1.90718114677456 0.460121816699439
1.93627356887304 0.471104503381133
1.96536599097152 0.481319840775023
1.99445841306999 0.492686202100619
2.02355083516847 0.503812766773776
2.05264325726695 0.514459738142056
2.08173567936543 0.524675075535946
2.11082810146391 0.535130209582274
2.13992052356239 0.545969018272505
2.16901294566087 0.556999664284687
2.19810536775935 0.570092561507841
2.22719778985783 0.581171166850511
2.25629021195631 0.592537528176107
2.28538263405479 0.605294710085847
2.31447505615327 0.616613112080955
2.34356747825175 0.62821927005899
2.37265990035023 0.638914200757758
2.40175232244871 0.649944846769939
2.43084474454719 0.661551004747974
2.45993716664566 0.671958179463815
2.48902958874414 0.682796988154045
2.51812201084262 0.694211308810129
2.5472144329411 0.704666442856458
2.57630685503958 0.714977698911323
2.60539927713806 0.724665483669848
2.63449169923654 0.733681837801544
2.66358412133502 0.742362476619826
2.6926765434335 0.750755359455182
2.72176896553198 0.758860486307611
2.75086138763046 0.766773775838089
2.77995380972894 0.774495228046617
2.80904623182742 0.781737086950266
2.8381386539259 0.789458539158793
2.86723107602438 0.796076926766102
2.89632349812286 0.802887151695362
2.92541592022133 0.809649417294135
2.95450834231981 0.81612392690998
2.98360076441829 0.822022924559973
3.01269318651677 0.827442328905089
3.04178560861525 0.832238261953863
3.07087803071373 0.837897462951417
3.09997045281221 0.842789314661167
3.12906287491069 0.847681166370918
3.15815529700917 0.852045465445303
3.18724771910765 0.856553642511151
3.21634014120613 0.860438348280658
3.24543256330461 0.865809793295286
3.27452498540309 0.870030214378207
3.30361740750157 0.874394513452592
3.33270982960005 0.878662893866002
3.36180225169852 0.881876169008681
3.390894673797 0.886048630761115
3.41998709589548 0.889165987242818
3.44907951799396 0.892571099707448
3.47817194009244 0.895784374850127
3.50726436219092 0.898997649992806
3.5363567842894 0.901875209822071
3.56544920638788 0.90465685099036
3.59454162848636 0.907390532828162
3.62363405058484 0.910699726631816
3.65272647268332 0.913481367800106
3.6818188947818 0.916694642942785
3.71091131688028 0.918948731475709
3.74000373897876 0.921730372643998
3.76909616107724 0.924176298498873
3.79818858317572 0.927053858328138
3.82728100527419 0.929020190878135
3.85637342737267 0.93146611673301
3.88546584947115 0.93338448995252
3.91455827156963 0.935494700493981
3.94365069366811 0.937700829696417
3.97274311576659 0.939858999568366
4.00183553786507 0.941921250779339
4.03092795996355 0.944319217303726
4.06002038206203 0.946813102489089
4.08911280416051 0.94849167905616
4.11820522625899 0.95041005227567
4.14729764835747 0.952184547503717
4.17639007045595 0.954198839384202
4.20548249255443 0.95573353795981
4.23457491465291 0.958275382475661
4.26366733675138 0.960241715025658
4.29275975884986 0.961776413601266
4.32185218094834 0.963454990168337
4.35094460304682 0.964989688743945
4.3800370251453 0.966428468658577
4.40912944724378 0.967963167234185
4.43822186934226 0.969258069157354
4.46731429144074 0.971080523715889
4.49640671353922 0.972231547647595
4.5254991356377 0.973958083545154
4.55459155773618 0.975540741451249
4.58368397983466 0.977123399357345
4.61277640193314 0.978562179271977
4.64186882403162 0.979905040525634
4.6709612461301 0.981151983118316
4.70005366822858 0.98254280370246
4.72914609032706 0.983166274998801
4.75823851242553 0.984221380269531
4.78733093452401 0.985228526209774
4.81642335662249 0.986427509471968
4.84551577872097 0.987338736751235
4.87460820081945 0.988345882691477
4.90370062291793 0.989448947292696
4.93279304501641 0.990360174571963
4.96188546711489 0.991127523859767
4.99097788921337 0.992038751139034
5.02007031131185 0.993045897079277
5.04916273341033 0.99362140904513
5.07825515550881 0.994053043019519
5.10734757760729 0.994820392307323
5.13643999970577 0.995299985612201
5.16553242180425 0.99573161958659
5.19462484390272 0.996259172221956
5.2237172660012 0.996642846865858
5.25280968809968 0.99702652150976
5.28190211019816 0.997362236823174
5.31099453229664 0.997841830128051
5.34008695439512 0.998129586110978
5.3691793764936 0.998465301424392
5.39827179859208 0.998609179415855
5.42736422069056 0.998801016737806
5.45645664278904 0.999040813390245
5.48554906488752 0.999280610042684
5.514641486986 0.999472447364635
5.54373390908448 0.999664284686586
5.57282633118296 0.999760203347561
5.60191875328144 0.999760203347561
5.63101117537992 0.999760203347561
5.6601035974784 0.999904081339024
5.68919601957687 0.999952040669512
5.71828844167535 0.999952040669512
5.74738086377383 0.999952040669512
5.77647328587231 0.999952040669512
5.80556570797079 0.999952040669512
5.83465813006927 1
};
\addlegendentry{$T_\mathrm{opt} \circ C_{\Theta, \mathrm{CIRA}}$}

\end{axis}

\end{tikzpicture}

%% file: dopplercc.bbl
\begin{thebibliography}{10}
\providecommand{\url}[1]{#1}
\csname url@samestyle\endcsname
\providecommand{\newblock}{\relax}
\providecommand{\bibinfo}[2]{#2}
\providecommand{\BIBentrySTDinterwordspacing}{\spaceskip=0pt\relax}
\providecommand{\BIBentryALTinterwordstretchfactor}{4}
\providecommand{\BIBentryALTinterwordspacing}{\spaceskip=\fontdimen2\font plus
\BIBentryALTinterwordstretchfactor\fontdimen3\font minus
  \fontdimen4\font\relax}
\providecommand{\BIBforeignlanguage}[2]{{%
\expandafter\ifx\csname l@#1\endcsname\relax
\typeout{** WARNING: IEEEtran.bst: No hyphenation pattern has been}%
\typeout{** loaded for the language `#1'. Using the pattern for}%
\typeout{** the default language instead.}%
\else
\language=\csname l@#1\endcsname
\fi
#2}}
\providecommand{\BIBdecl}{\relax}
\BIBdecl

\bibitem{studer_cc}
C.~Studer, S.~Medjkouh, E.~G{\"{o}}n{\"{u}}ltas, T.~Goldstein, and
  O.~Tirkkonen, ``{Channel Charting: Locating Users within the Radio
  Environment using Channel State Information},'' \emph{CoRR}, vol.
  abs/1807.05247, 2018.

\bibitem{fraunhofer_cc}
M.~Stahlke, G.~Yammine, T.~Feigl, B.~M. Eskofier, and C.~Mutschler, ``{Indoor
  Localization with Robust Global Channel Charting: A Time-Distance-Based
  Approach},'' \emph{IEEE Transactions on Machine Learning in Communications
  and Networking}, 2023.

\bibitem{stephan2023angle}
P.~Stephan, F.~Euchner, and S.~ten Brink, ``{Angle-Delay Profile-Based and
  Timestamp-Aided Dissimilarity Metrics for Channel Charting},'' \emph{IEEE
  Transactions on Communications}, 2024.

\bibitem{pihlajasalo2020absolute}
J.~Pihlajasalo, M.~Koivisto, J.~Talvitie, S.~Ali-L{\"o}ytty, and M.~Valkama,
  ``{Absolute Positioning with Unsupervised Multipoint Channel Charting for 5G
  Networks},'' in \emph{2020 IEEE 92nd Vehicular Technology Conference
  (VTC2020-Fall)}.\hskip 1em plus 0.5em minus 0.4em\relax IEEE, 2020.

\bibitem{taner2023channel}
S.~Taner, V.~Palhares, and C.~Studer, ``{Channel Charting in Real-World
  Coordinates},'' in \emph{2023 IEEE Global Communications Conference}, 2023.

\bibitem{asilomar2023}
F.~Euchner, P.~Stephan, and S.~t. Brink, ``{Augmenting Channel Charting with
  Classical Wireless Source Localization Techniques},'' in \emph{57th Asilomar
  Conference}, 2023.

\bibitem{chan}
Y.-T. Chan and F.~L. Jardine, ``{Target localization and tracking from
  Doppler-shift measurements},'' \emph{IEEE Journal of Oceanic Engineering},
  vol.~15, no.~3, pp. 251--257, 1990.

\bibitem{shames}
I.~Shames, A.~N. Bishop, M.~Smith, and B.~D.~O. Anderson, ``{Doppler Shift
  Target Localization},'' \emph{IEEE Transactions on Aerospace and Electronic
  Systems}, vol.~49, no.~1, pp. 266--276, 2013.

\bibitem{stahlke2023velocity}
M.~Stahlke, G.~Yammine, T.~Feigl, B.~M. Eskofier, and C.~Mutschler,
  ``{Velocity-Based Channel Charting with Spatial Distribution Map Matching},''
  \emph{arXiv preprint arXiv:2311.08016}, 2023.

\bibitem{taner2023streaming}
S.~Taner, M.~Guillaud, O.~Tirkkonen, and C.~Studer, ``{Channel Charting for
  Streaming CSI Data},'' in \emph{{2023 57th Asilomar Conference}}, 2023.

\bibitem{dichasus2021}
F.~Euchner, M.~Gauger, S.~D\"orner, and S.~ten Brink, ``{A Distributed Massive
  MIMO Channel Sounder for "Big CSI Data"-driven Machine Learning},'' in
  \emph{25th ITG Workshop on Smart Antennas}, 2021.

\bibitem{euchner2022geometry}
F.~Euchner, P.~Stephan, M.~Gauger, and S.~ten Brink, ``{Geometry-Based Phase
  and Time Synchronization for Multi-Antenna Channel Measurements},'' in
  \emph{2022 IEEE Globecom Workshops}.\hskip 1em plus 0.5em minus 0.4em\relax
  IEEE, 2022.

\bibitem{brown2005method}
D.~R. Brown, G.~B. Prince, and J.~A. McNeill, ``{A method for carrier frequency
  and phase synchronization of two autonomous cooperative transmitters},'' in
  \emph{IEEE 6th Workshop on Signal Processing Advances in Wireless
  Communications, 2005.}\hskip 1em plus 0.5em minus 0.4em\relax IEEE, 2005, pp.
  260--264.

\bibitem{zhao2024signature}
L.~Zhao, Y.~Yang, Q.~Xiong, H.~Wang, B.~Yu, F.~Sun, and C.~Sun, ``{A Signature
  Based Approach Towards Global Channel Charting with Ultra Low Complexity},''
  \emph{arXiv preprint arXiv:2403.20091}, 2024.

\end{thebibliography}
